\documentclass{aa}  

\usepackage{graphicx}
\usepackage{txfonts}

\usepackage[breaklinks,colorlinks,citecolor=blue,linkcolor=blue,urlcolor=blue]{hyperref}
\begin{document}

   \title{The VMC survey}
   
   \subtitle{XXVIII. Improved measurements of the proper motion of the Galactic globular cluster 47~Tucanae\thanks{Based on observations made with VISTA at the Paranal Observatory under program ID 179.B-2003}}

   \author{Florian Niederhofer
          \inst{1}
          \and
          Maria-Rosa L. Cioni
          \inst{1,2}
          \and
          Stefano Rubele
          \inst{3,4}
          \and
          Thomas Schmidt
          \inst{1}
          \and
          Kenji Bekki
          \inst{5}
          \and
          Richard de Grijs
          \inst{6,7}
          \and
          Jim Emerson
          \inst{8}
          \and
          Valentin D. Ivanov
          \inst{9,10}
          \and
          Joana M. Oliveira
          \inst{11}
          \and
          Monika G. Petr-Gotzens
          \inst{10}
          \and
          Vincenzo Ripepi
          \inst{12}
          \and
          Ning-Chen~Sun
          \inst{6}
          \and
          Jacco~Th.~van~Loon
          \inst{11}
          }

   \institute{Leibniz-Institut f\"ur Astrophysik Potsdam, An der Sternwarte 16, 14482 Potsdam, Germany\\
              \email{fniederhofer@aip.de}
              \and
              University of Hertfordshire, Physics Astronomy and Mathematics, College Lane, Hatfield AL10 9AB, UK
              \and
              Osservatorio Astronomico di Padova $-$ INAF, Vicolo dell'Osservatorio 5, I-35122 Padova, Italy
              \and
              Dipartimento di Fisica e Astronomia, Universita di Padova, Vicolo dell’Osservatorio 2, I-35122 Padova, Italy
              \and
              ICRAR, M468, University of Western Australia, 35 Stirling Hwy, 6009 Crawley, Western Australia, Australia
              \and
              Kavli Institute for Astronomy \& Astrophysics and Department of Astronomy, Peking University, Yi He Yuan Lu 5, Hai Dian District, Beijing 100871, China
              \and
              International Space Science Institute--Beijing, 1 Nanertiao, Zhongguancun, Hai Dian District, Beijing 100190, China
              \and
              Astronomy Unit, School of Physics and Astronomy, Queen Mary University of London, Mile End Road, London E1 4NS, UK
              \and
              European Southern Observatory, Ave. Alonso de Cordova 3107, Vitacura, Santiago, Chile
              \and
              European Southern Observatory, Karl-Schwarzschild-Str. 2, D-85748 Garching bei M\"{u}nchen, Germany
             \and
             Lennard-Jones Laboratories, 
             Keele University, ST5 5BG, UK
              \and
              INAF-Osservatorio Astronomico di Capodimonte, via Moiariello 16, 80131, Naples, Italy
             }

   \date{Received; accepted}

 
  \abstract
  {We use deep multi-epoch point-spread function (PSF) photometry taken with the Visible and Infrared Survey Telescope for Astronomy (VISTA) to measure and analyze the proper motions of stars within the Galactic globular cluster 47~Tucanae (47 Tuc, NGC~104). The observations are part of the ongoing near-infrared VISTA survey of the Magellanic Cloud system (VMC).  The data analyzed in this study correspond to one VMC tile which covers a total sky area of 1.77 deg$^2$.
Absolute proper motions with respect to $\sim$9,070 background galaxies are calculated from a linear regression model applied to the positions of stars in 11 epochs in the $K_s$ filter. The data extend over 
a total time baseline of about 17 months. We found an overall median proper motion of the stars within 47~Tuc of 
$(\mu_{\alpha}\mathrm{cos}(\delta),~\mu_{\delta}) = (+5.89 \pm 0.02~\mathrm{(statistical)} \pm 0.13~\mathrm{(systematic)}, -2.14 \pm 0.02~\mathrm{(statistical)} \pm 0.08~\mathrm{(systematic)})$~mas~yr$^{-1}$, based on the measurements of $\sim$35,000 individual sources between 5$\arcmin$ and 42$\arcmin$ from the cluster center. We compared our result to the proper motions from the newest US Naval Observatory CCD Astrograph Catalog (UCAC5)  which includes data from the \textit{Gaia} data release 1. Selecting cluster members ($\sim$2,700 stars) we found a median proper motion of
$(\mu_{\alpha}\mathrm{cos}(\delta),~\mu_{\delta}) = (+5.30 \pm 0.03~\mathrm{(statistical)} \pm 0.70~\mathrm{(systematic)}, -2.70 \pm 0.03~\mathrm{(statistical)} \pm 0.70~\mathrm{(systematic)})$~mas~yr$^{-1}$.
Comparing the results with measurements in the literature, we found that the values derived from the VMC data are consistent with the UCAC5 result, and are close to measurements obtained using the \textit{Hubble Space Telescope}. 
We combined our proper motion results with radial velocity measurements from the literature and reconstructed the orbit of 47~Tuc, finding that the cluster is on an orbit with a low ellipticity and is confined within the inner $\sim$7.5~kpc of the Galaxy. 
We show that the use of an increased time baseline in combination with PSF-determined stellar centroids in crowded regions significantly improves the accuracy of the method. In future works, we will apply the methods described here to more VMC tiles to study in detail the kinematics of the Magellanic Clouds.
}


   \maketitle
%

\section{Introduction \label{sec:intro}}

The VISTA survey of the Magellanic Clouds (VMC, \citealt{Cioni11}) is a multi-epoch near-infrared survey mostly designed to study in detail the three-dimensional structure, the overall and internal kinematics, as well as the spatially resolved star formation history of the two Milky Way satellites, the Large and the Small Magellanic Cloud (LMC and SMC). The Magellanic Clouds are currently interacting both with each other and with the Milky Way and form a composite system including a Bridge connecting the two Clouds and a Stream \citep[see e.g.][for a review]{D'Onghia16}. Especially the SMC is revealed to have a complex structure \citep[e.g.][]{Subramanian12, Ripepi17, Subramanian17}. 

By virtue of the multi-epoch nature of the VMC survey, its data can be used for dynamical  studies, e.g. proper motion measurements of resolved stars. In a pilot study, \citet{Cioni14} determined the median motion of the LMC in the plane of the sky both by combining one VMC tile with data from the Two Micron All Sky Survey (2MASS) and from VMC data alone. The combination of VMC with 2MASS gives a time baseline of about 10~years whereas the VMC observations alone span about 1~year. Later, \citet{Cioni16} calculated the proper motion of various stellar populations of the SMC from a tile which also includes the Galactic globular cluster 47~Tucanae (47~Tuc, NGC~104), using a time baseline of about 12~months.

Proper motion measurements of stellar populations are a key ingredient to understand their evolution and origins. A vast number of works have already reported the proper motions of different extragalactic stellar systems, including nearby dwarf galaxies \citep[e.g.][]{Sohn13, Sohn17, Massari13}, the Sagittarius Stream \citep{Sohn15} and the Magellanic Clouds \citep[e.g.][]{Kallivayalil06, Costa09,Vieira10,Costa11,Kallivayalil13,Cioni14,Cioni16}. Of special interest are also Galactic globular clusters \citep[e.g.][]{AndersonKing03, Bedin03, Milone06, Anderson10, Bellini10, Bellini14, Cadelano17, Sariya17}. 
Accurate stellar motion measurements can shed light on the internal kinematics of the clusters probing, for example, the rotation of the cluster in the plane of the sky \citep[e.g.][]{AndersonKing03,Bellini17} and the presence of tidal tails. 
Combining the proper motions with spectroscopic radial velocities 
of a cluster one can obtain the full three-dimensional velocity 
vector of a cluster and the cluster's orbit within the Milky Way
\citep[e.g.][]{Casetti-Dinescu07,Casetti-Dinescu10,Casetti-Dinescu13,Cadelano17, Koch17}. This allows one to probe 
the kinematic structure of different components of our Galaxy.
Stellar proper motions within a single cluster can also be used as a criterion to separate cluster members from unrelated field stars, e.g. for photometric studies of multiple populations within a cluster \citep[see e.g.][]{Piotto12, Richer13, Milone15, Milone17} or for follow up spectroscopic studies. By comparing the line-of-sight velocity dispersion with the one in the plane of the sky, proper motion measurements can also be used for an independent estimate of the distance of globular clusters \citep[see e.g.][]{McLaughlin06, Watkins17} or the distance to the nuclear star cluster in the Milky Way \citep{Chatzopoulos15}. The latter measurement can be used to obtain the distance of the Sun from the Galactic Center.

Soon, the \textit{Gaia} space mission will provide all-sky high-quality astrometric and proper motion measurements for sources brighter than $\sim$20~mag in the $G$ filter. In September 2016, first data from \textit{Gaia} were released (\textit{Gaia} data release~1, DR1, \citealt{Gaia16}). Crossmatching the available data from \textit{Gaia} DR1 with the \textit{Hipparcos} Tycho2 stellar catalog \citep{Hog2000}, accurate proper motions can already be calculated for stars in common thanks to the extended time baseline of $\sim$25 years \citep{Michalik15, Lindegren16}. This is referred to as the Tycho-\textit{Gaia} Astrometric Solution (TGAS). Data from the TGAS catalog have already been used to measure galactic rotations within the Magellanic Clouds \citep{vanderMarel16} and the absolute proper motions of a sample of five Galactic globular clusters, including 47~Tuc \citep{Watkins17}. 

As a preparatory work for future studies, we re-calculate the proper motions of the stars within the globular cluster 47~Tuc which are presented in \citet{Cioni16} using data from the VMC survey. \citet{Cioni16} used for their calculation a total time interval of 12 months and stellar positions from the VISTA Data Flow System (VDFS) pipeline catalog. There, the stellar centroids were determined as the intensity-weighted isophotal center-of-mass in the $x$ and $y$ directions.
Here we will use improved stellar centroid determinations derived from point-spread function (PSF) photometry in combination with an increased time baseline of 17 months. By comparing the updated results with values from the literature we show that the use of PSF photometry and a longer time baseline significantly improves the accuracy of the proper motion determination, with respect to previous results obtained using the VMC data, especially in crowded regions. In upcoming works, we will apply the techniques described in this paper to the VMC tiles covering the SMC to study the over all motion and internal kinematics within the galaxy.

The paper is structured as follows: In Section~\ref{sec:obs}, we describe the observational data sets used, the reduction procedure, and the construction of the data catalogs. We explain our methods to calculate the proper motions in Section~\ref{sec:PM} and present the results and the analysis in Section~\ref{sec:results}. Concluding remarks and further prospects are given in Section~\ref{sec:conclusions}.

\section{Observations and Photometry \label{sec:obs}}

In this study, we use multi-epoch observations taken with the Visible and Infrared Survey Telescope for Astronomy (VISTA, \citealt{Emerson10}) in the $K_s$ filter (central wavelength 2.15~$\mu$m). These observations are part of the VMC survey. The VISTA infrared camera VIRCAM \citep{Emerson06, Dalton06} consists of 16 detectors, covering 2048 $\times$ 2048 pixels each. The pixel size is 0$\farcs$34. The detectors are arranged in a 4 $\times$ 4 array with gaps between the detectors that are 90\% in the $x$ direction and 42.5\% of the width of the chip in the $y$ direction. In order to observe a contiguous area on the sky, six individual exposures (pawprints) with specific offsets to cover the gaps between the detectors are combined to form a single tile image. Therefore, a single source can be observed up to six times in an individual tile, depending on its position in the overlap regions of the pawprints. The final tile covers an area of $\sim$1.77~deg$^2$. We refer the interested reader to \citet{Cioni11} for a detailed description of the VMC survey and the observing strategy. 

47~Tuc is located in close projected vicinity to the SMC and falls within VMC tile SMC~5\_2 (centered at $\alpha_{2000}=00^{\mathrm{h}}26^{\mathrm{m}}41\overset{\mathrm{s}}.69$ and $\delta_{2000}=-71\degr56\arcmin35\farcs88$). The tile has three epochs of observations in the $J$ and $Y$ filters and 12 epochs in the $K_s$ filter. Eleven of these epochs in $K_s$ are deep observations with exposure times of 750~s each whereas the remaining one is split into two shallower observations of 375~s each. For the proper motion studies, we will use the 11 $K_s$ epochs with long exposures and the first short exposure, to get a longer time baseline. We do not include the second short exposure since it was observed on the same night as the first deep exposure. The first epoch of observation was performed on 9 June 2011 and the last one on 10 November 2012 which gives us an overall time baseline of about 17 months. Table~\ref{tab:multi-epoch} summarizes the details of the individual observations.

\begin{table} \small
\centering
\caption{Observing log for all SMC 5\_2 observations in $K_s$. The epochs denoted TK are the deep observations. The epochs with shallow observations are denoted CK (see text).  \label{tab:multi-epoch}}
\begin{tabular}{@{}l@{ }c@{ }c@{ }c@{ }c@{ }c@{}}
\hline\hline
\noalign{\smallskip}
Date of  & ~Identifier~ & ~Seeing~ & ~Limiting~ & ~Airmass~\\
Observation~ & & (arcsec) & ~Magnitude per Tile\tablefootmark{a}~ & \\
\noalign{\smallskip}
\hline
\noalign{\smallskip}

09-06-2011                              & CK2   & 0.93 &  18.82 &  1.584\\
16-11-2011\tablefootmark{b}   & CK1   & 0.98 &  19.02 &  1.512\\
16-11-2011                              &  TK1   & 0.86 &  19.39 &  1.553\\
20-11-2011                              &  TK2   & 0.92 &  19.41 &  1.480\\
25-11-2011                              &  TK3   & 0.96 &  19.41 &  1.479\\
06-12-2011                              &  TK4   & 0.76 &  19.34 &  1.512\\
13-06-2012\tablefootmark{b}  &  TK5   & 0.96 &  18.82 &  1.660\\
25-07-2012                             &  TK6   & 0.85 &  19.47 &  1.490\\
14-08-2012                             &  TK7   & 0.92 &  19.47 &  1.500\\
05-09-2012                             &  TK8   & 1.00 &  19.35 &  1.485\\
22-09-2012                             &  TK9   & 0.79 &  19.48 &  1.492\\ 
10-10-2012                             &  TK10 & 0.93 &  19.58 &  1.549\\
10-11-2012                             &  TK11  & 0.90 &  19.49 &  1.547\\

\noalign{\smallskip}
\hline
\end{tabular}
\tablefoot{
\tablefoottext{a}{Sources with a signal to noise ratio $\geq$10}\\
\tablefoottext{b}{Epoch not used in proper motion calculation}\\
}
\end{table}

For the proper motion calculations we made use of the PSF photometry that was performed on each pawprint for every epoch separately (see \citealt{Rubele15}, for more details). The images were reduced by the Cambridge Astronomy Survey Unit (CASU) with the VDFS pipeline version 1.3 (\citealt{Irwin04} and \url{http://casu.ast.cam.ac.uk/surveys-projects/vista/data-processing/version-log}) and retrieved from the VISTA Science Archive\footnote{\url{http://horus.roe.ac.uk/vsa}} (VSA, \citealt{Cross12}).
The pipeline deals with linearity,  dark current, flat fielding, large-scale sky background emission, etc. It also shifts and combines the jitter series to a single deep stack image, i.e. the pawprint image. 
We derived a PSF model for each pawprint, using IRAF/DAOPHOT\footnote{IRAF is distributed by the National Optical Astronomy Observatory which is operated by the Association of Universities for Research in Astronomy (AURA) under a cooperative agreement with the US National Science Foundation.} tasks, assuming a linearly variable PSF model over each detector. 
We performed the photometry on the 
images using the ALLSTAR \citep{Stetson87} task. This routine calculates refined centers of the stars, their magnitudes and local sky values by iteratively fitting model PSFs to the stars and then subtracting them from the image. For final aperture corrections and absolute photometric calibration we used as references the VSA data releases and the 2MASS data as described by \citet{Rubele15}.

For our photometric analysis we also made use of a multi-band
deep tile catalog combining the multiple epochs. We applied the method described in \citet{Rubele15} to homogenize the PSF in the various pawprints to a constant reference PSF model. The pawprints from all epochs, which now have a constant PSF, were combined into a single deep image and the PSF photometry was performed on this image accordingly. The final catalog including all three filters was created by correlating the single-band catalogs with a matching radius of 1~arcsec choosing the nearest neighbor in case of multiple matches. 

\begin{figure}
\centering
\includegraphics[width=\columnwidth]{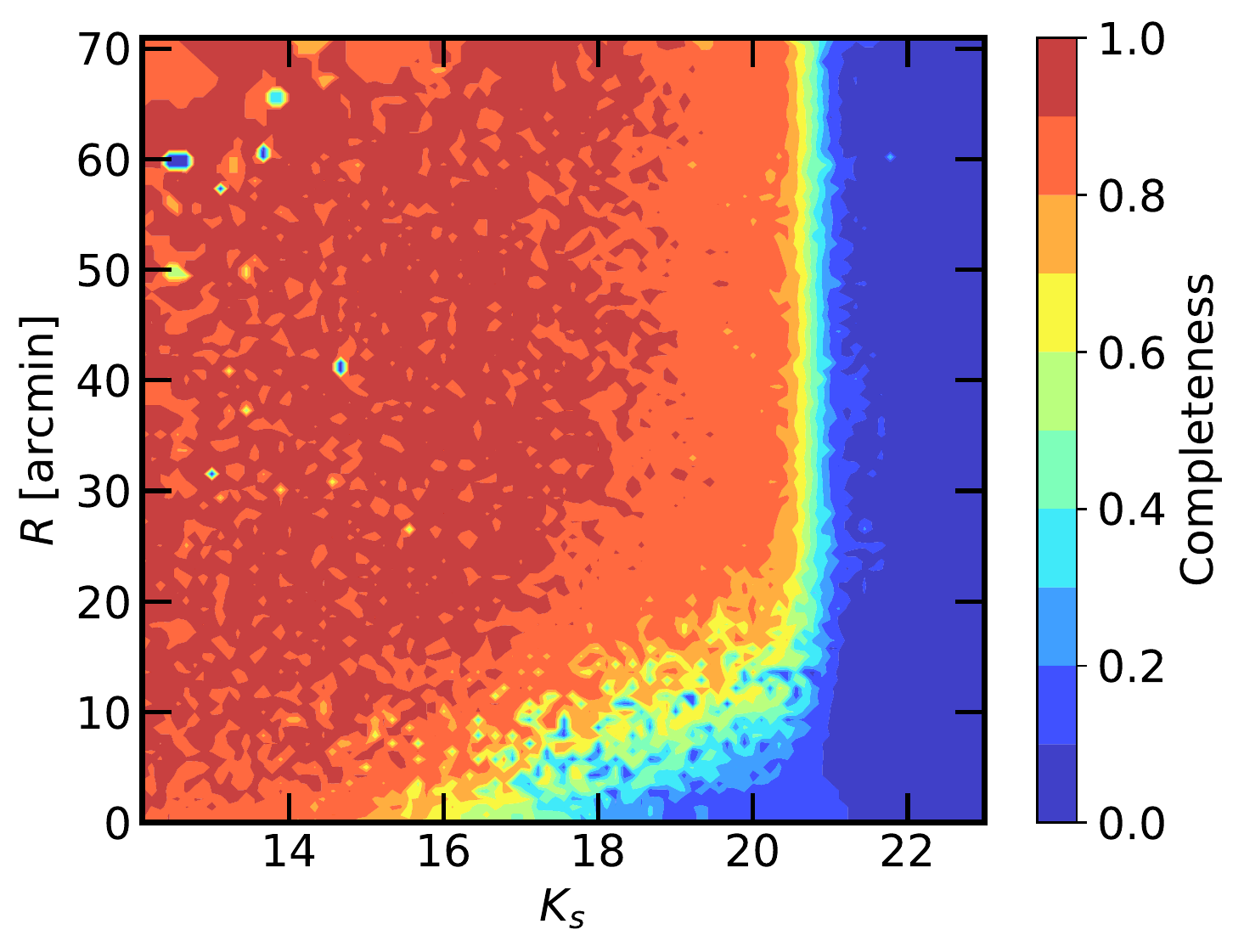}
\caption{Two-dimensional completeness map as a function of $K_s$ magnitude and the distance $R$ from the center of 47~Tuc.}
\label{fig:completeness}
\end{figure}

We assessed the completeness of the photometry by means of extensive artificial star tests. Figure \ref{fig:completeness} shows a two-dimensional map of the local completeness, as a function of $K_s$ magnitude and the distance to the center of 47~Tuc. We can see that the 50\% completeness limit is at $\sim$20.7~mag for regions as close as about 15 to 20~arcmin from the cluster center. This limit drops to about 16~mag in the innermost parts of the cluster. 

\begin{figure}
\centering
\includegraphics[width=\columnwidth]{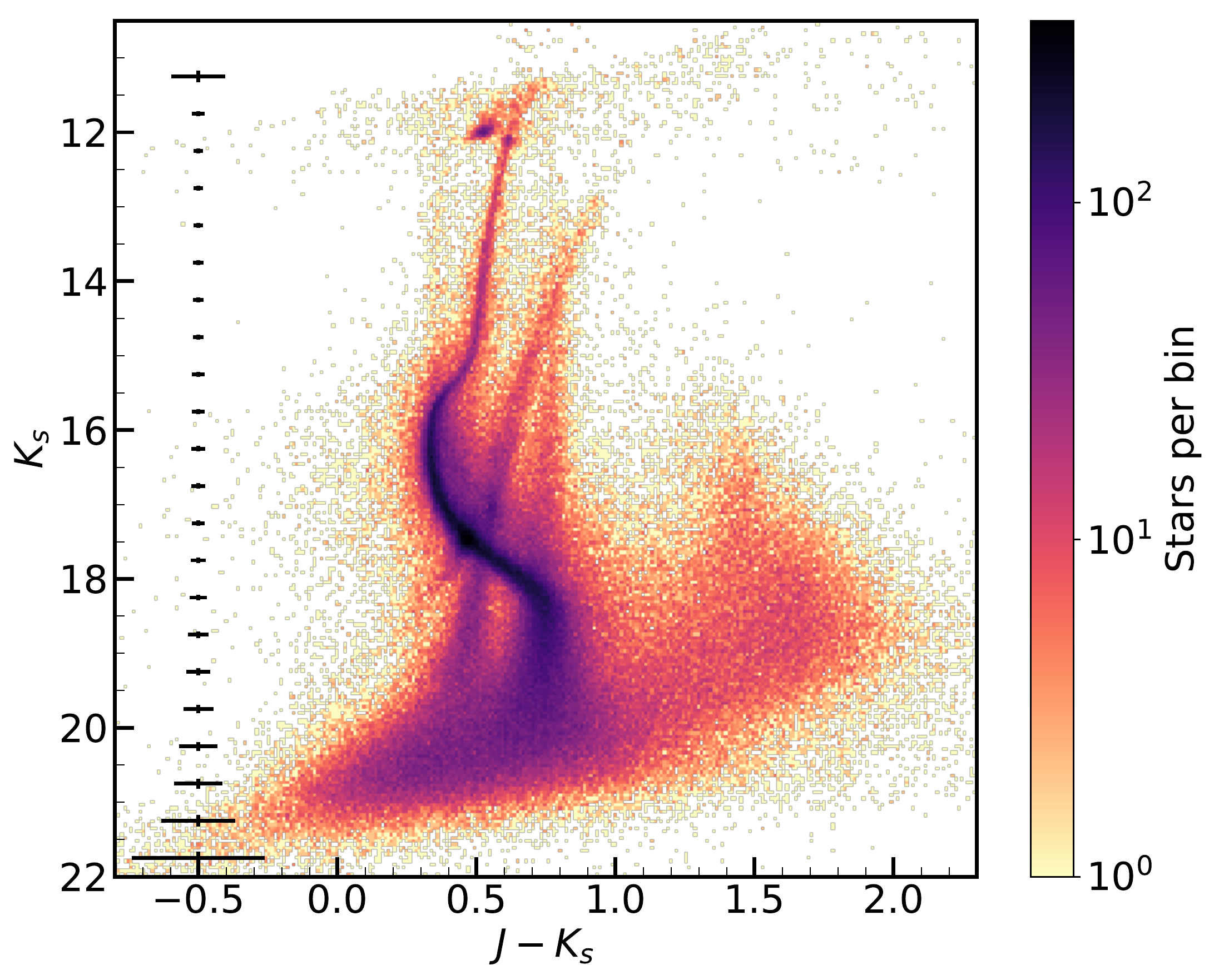}
\caption{Hess diagram in the $K_s$ vs $J-K_s$ color-magnitude space of all sources detected in tile SMC 5\_2 based on PSF photometry. The mean photometric errors as a function of the $K_s$ magnitude are indicated on the left-hand side as black bars. Note that the color scale is in logarithmic units.}
\label{fig:hess}
\end{figure}

Figure \ref{fig:hess} shows a stellar density plot (Hess diagram) in the $K_s$ vs $J-K_s$ color-magnitude space of all sources detected in $J$ and $K_s$ within tile SMC 5\_2 (see also Figure~3 in \citealt{Cioni16}). Also indicated as black bars on the left-hand side are the mean photometric uncertainties as a function of $K_s$ magnitude. The main sequence (MS) of 47~Tuc is readily recognizable as the track of the highest stellar density. The main sequence turn-off (MSTO) is at a $K_s$ magnitude of $\sim$15.5 and is followed by 47~Tuc's red giant branch (RGB), which extends up to $K_s\sim$11.5~mag where stars start to saturate (because of different software packages, the stars saturate at fainter magnitudes than in the standard VDFS aperture photometry catalogs). The cluster's red clump (RC) can be seen as an enhancement in stellar density at $J-K_s\sim0.5$~mag and $K_s\sim$12~mag. The evolved stars of the SMC form the diagonal sequence that crosses the MS of 47~Tuc, whereas the galaxy's RC lies almost exactly on top of the cluster's MS at $J-K_s\sim0.4$~mag and $K_s\sim17.5$~mag. The nearly vertical sequence at $J-K_s\sim0.75$~mag is composed of Milky Way foreground stars. Finally, the triangularly shaped structure at $J-K_s$ colors $>$1.0~mag is mostly populated by background galaxies. 

\begin{figure}
\centering
\includegraphics[width=\columnwidth]{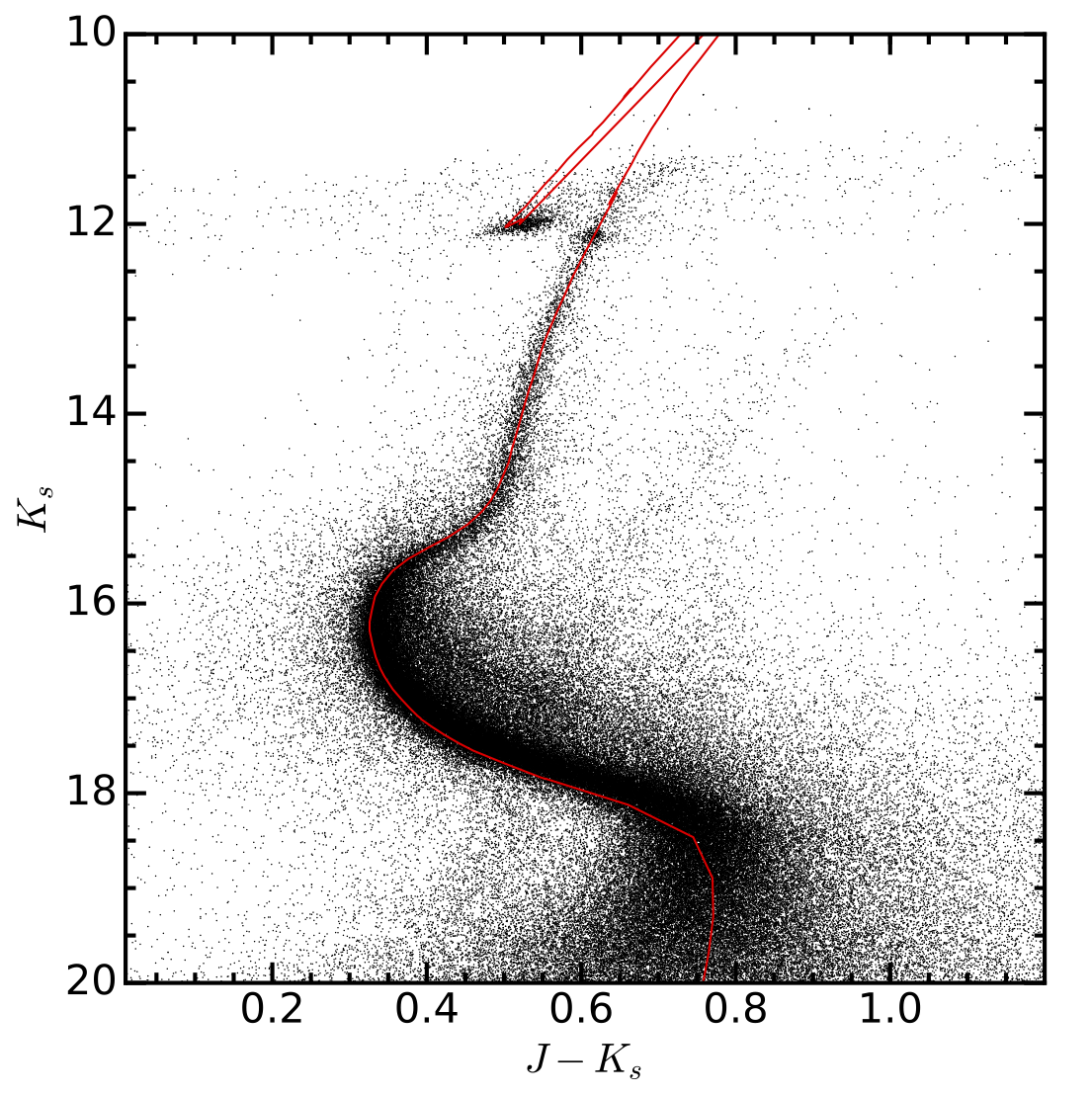}
\caption{CMD of all stars within 20$\arcmin$ from the center of 47 Tuc. 
The diagram is dominated by stars belonging to the cluster, but the sequences of the Milky Way and the SMC are still visible.
Also shown as a red line is an isochrone at 11.8~Gyr from the PARSEC model set \citep{Bressan12}, with Z = 0.0033, $E(B-V) = 0.04$~mag and $(m-M)_0$ = 13.30~mag.}
\label{fig:cmd}
\end{figure}

In this paper, we concentrate our analysis on the globular cluster 47~Tuc. A color-magnitude diagram (CMD) of this cluster using the VMC data was first shown and discussed by \citet{Li14}. Figure \ref{fig:cmd} shows a $K_s$ vs $J-K_s$ CMD of stars that are within 20$\arcmin$ of the center of 47~Tuc (we adopt the cluster center coordinates as determined by \citealt{Li14}: $\alpha_{2000} = 00^{\mathrm{h}}24^{\mathrm{m}}04\overset{\mathrm{s}}.80$, $\delta_{2000} = -72\degr04\arcmin48\arcsec$). This CMD is dominated by stars belonging to the cluster. However, the SMC population is still visible here. Also shown as a red line is an isochrone from the PARSEC evolutionary models\footnote{\url{http://stev.oapd.inaf.it/cgi-bin/cmd}} \citep{Bressan12} following the sequences of the cluster. For the isochrone, we adopted a metallicity of Z = 0.0033 ([Fe/H] = $-$0.76~dex), consistent with various measurements in the literature (e.g., \citealt{Harris96}, 2010 edition; \citealt{McLaughlin05, Carretta09, O'Malley17}) and an age of 11.8~Gyr (log(age/yr) = 10.07) which is in agreement with recent measurements \citep[e.g.][]{Brogaard17, O'Malley17}. Additionally, to match the observations, we have to determine the interstellar extinction and the distance to the cluster. For the extinction, we adopted the extinction coefficients in the various VISTA filters as given in \citet{Rubele15} that are calculated assuming an extinction law with $R_V$ = 3.1 \citep{Cardelli89}. These coefficients are $A_J$ = 0.283$A_V$ and $A_{K_s}$ = 0.114$A_V$. We found  $A_{K_s}$ = 0.014~mag, which gives a corresponding extinction $A_V$ = 0.124~mag and color excess $E(B-V)$ = 0.04~mag. 
For the distance modulus we found  $(m-M)_0$ = 13.30~mag, which corresponds to a distance of 4.57~kpc. This is in agreement with the value of 4.6$\pm$0.2 obtained by \citet{McDonald11} who fitted an isochrone to the Hertzsprung-Russell diagram and also close to the average of recent distance measurements ($4.53^{+0.08}_{-0.04}$~kpc, \citealt{Bogdanov16}). \citet{Li14} used the $Y$ and $K_s$ filter combination in their analysis of 47~Tuc. They assumed a higher metallicity of [Fe/H] = $-$0.55~dex and also an older age of 12.5~Gyr. Using these values, they found a color excess $E(B-V)$ of 0.04~mag and a distance modulus $(m-M)_0$ of 13.40~mag.

\section{Proper Motion Calculations \label{sec:PM}}

Our strategy to determine the proper motions of the stars in tile SMC 5\_2 was to calculate them separately for each detector and each pointing. This way we minimize systematic effects that may arise when combining several detectors or pointings. For each of the 12 epochs there are 96 individual catalogs (16 detectors, 6 pointings) resulting in 1152 catalogs in total. These catalogs contain, besides the name of the observation and the detector number, for each star the RA and Dec coordinates, the magnitude and its uncertainty, the sharpness and the $x$ and $y$ coordinates on the detector chip. In the following, we will outline the different steps to calculate the proper motions of the stars. The overall procedure is similar to that described by \citet{Cioni16}.

\subsection{Star and Galaxy Catalogs \label{catalogs}}

In a first step we selected from the deep multi-band catalog only the sources that have detections in the $J$ and $K_s$ filters and assigned each source a unique identification number to be able to trace them throughout the calculations. We then cross-correlated this source list with the single-epoch $K_s$ catalogs, using the IRAF task \textit{xyxymatch} with a matching radius of 0.5~arcsec. This way we removed spurious detections from the calculations. We also assigned each catalog the Mean Julian Date (MJD) of the respective observation which was extracted from the multi-epoch aperture photometry files queried from the VSA website, which have been used by \citet{Cioni16}.

\begin{figure}
\centering
\includegraphics[width=0.9\columnwidth]{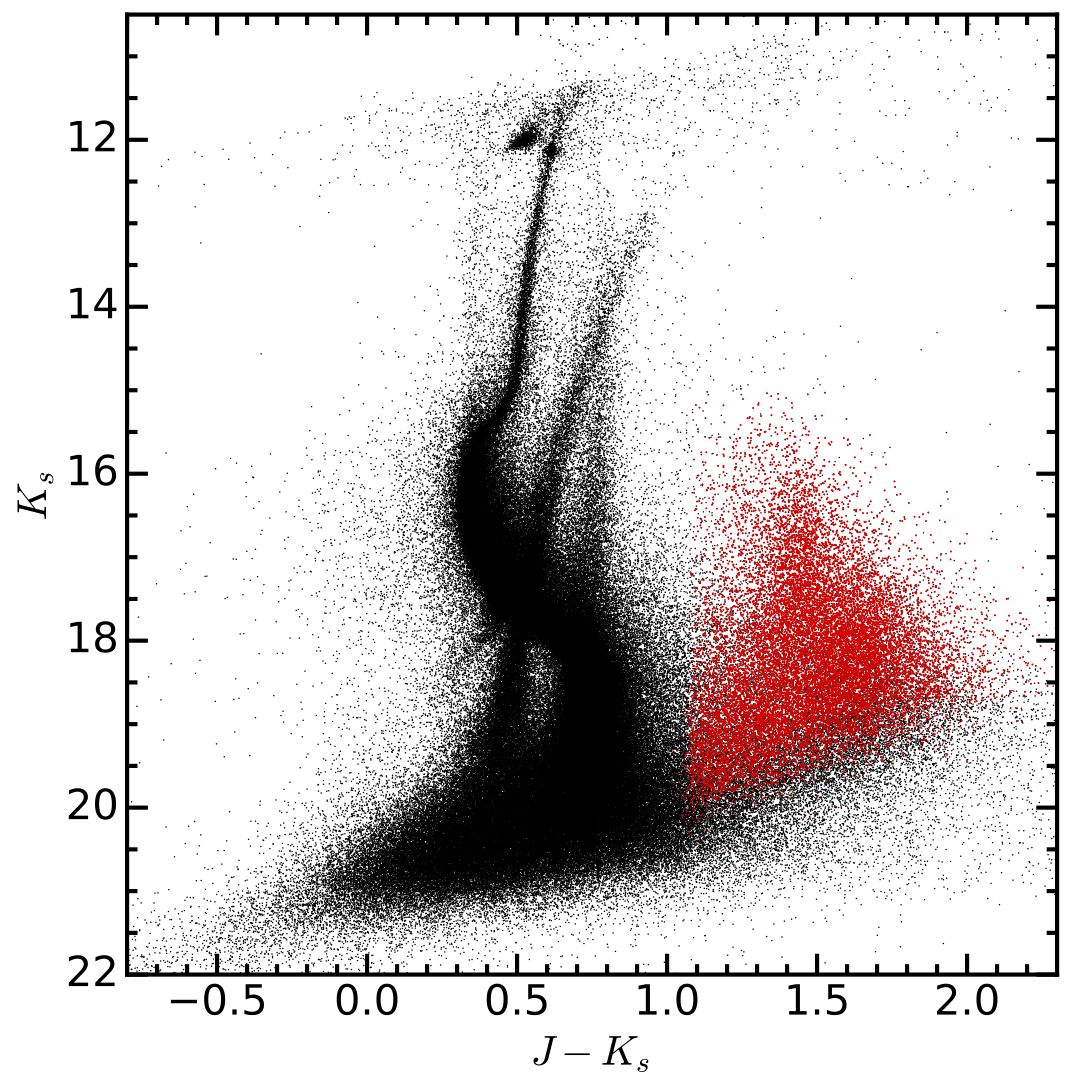}
\caption{$K_s$ vs $J-K_s$ CMD of unique sources detected in tile SMC 5\_2. The final selected sample of background galaxies used for the transformations is colored red.}
\label{fig:galaxies_cmd}
\end{figure}

Besides the different stellar populations from the cluster itself, the Milky Way and the SMC, the catalogs also contain detections of background galaxies (see Section~\ref{sec:obs} and Figure~\ref{fig:hess}). 
The reflex proper motion of these distant background objects is assumed to be zero. 
We will use the galaxies in a next step for an initial transformation of the various epochs to a common frame of reference. An additional, refined transformation will also be performed using stars of the cluster itself.
Based on several selection criteria, which we discuss below, we identified sources from the catalogs that are most likely galaxies and split our catalogs into two separate sets, one containing only stars and the other containing only the galaxies. For each source in the deep multi-band catalog there is a discrete stellar probability parameter which can have four values (0.0, 0.33, 0.67 and 1.0). It is calculated from the position in the color-color diagram, the local completeness and the sharpness. As background galaxies we selected sources that are classified as stars with a probability of less than 34\%, and also have $J-K_s$ colors greater than 1.0, as well as a sharpness parameter larger than 0.3 in both filters. At a sharpness of $\sim$0.3 the sequence of the galaxies separates from the one of the stars in a sharpness vs magnitude diagram. 
Blends of dwarf stars that could be miss-classified as galaxies due to their shapes are expected to be bluer than $J-K_s=1.0$ and should not contaminate our galaxy sample by much.
To obtain only well measured galaxies, we additionally restricted our final sample to galaxies with photometric errors $<$0.1~mag and $K_s$ magnitudes $>$15~mag, also to avoid contamination by saturated stars. 
Since we set a limit on the photometric error of the galaxies, we do not employ a faint limit in our galaxy selection. Using a combination of selection criteria in combination with conservative limits, we make sure that we get a sample of galaxies that is as clean as possible. The only potential contaminants are young stellar objects in the SMC that are expected to fall into the chosen color range and have extended emission due to circumstellar material. But the fraction of these objects contributing to our galaxy sample is negligible (a few per cent or less) given the size of the field and the location at the outskirts of the SMC. 
We ended up with a list of about 18,100 galaxies (see Figure~\ref{fig:galaxies_cmd} for a CMD illustrating the selected galaxies). 

\subsection{The Common Reference Frame\label{sec:reference}}

In the next step we transformed the catalogs from the various epochs to a common reference frame since there might be some small offsets and/or rotations between observations at different epochs. 
We performed the transformation in two steps. First, we used the previously extracted galaxies, which are assumed to represent a fixed grid of non-moving points, to perform an initial transformation. Then we used stars of 47~Tuc itself for a second, refined transformation. The first step using galaxies is necessary since large initial shifts between single epochs can lead to spurious matches in the inner, crowed regions of the cluster and reduce the quality of the transformation.
For the cluster stars we selected only well-measured stars (photometric errors $\sigma(K_s)\leq0.05$~mag, since we used only the $K_s$ filter for the proper motion determinations) from our catalog. To get a clean sample of cluster stars, we minimized contamination by the other stellar populations present in tile SMC 5\_2 which are unrelated to the cluster itself (see e.g. Figure~\ref{fig:hess}). Especially at fainter magnitudes these populations overlap significantly with the 47~Tuc stars in the CMD. We therefore selected only those stars in the CMD that are brighter than the RC feature of the SMC, which lies on top of the MS of 47~Tuc at $K_s\sim17$~mag. Additionally, we applied a mask in the color-magnitude space that follows the main CMD features of 47~Tuc (see Figure~\ref{fig:mask}). Our final list of cluster stars for the definition of the reference frame contains about 46,200 objects.

\begin{figure}
 \begin{tabular}{c}
  \includegraphics[width=7.0cm]{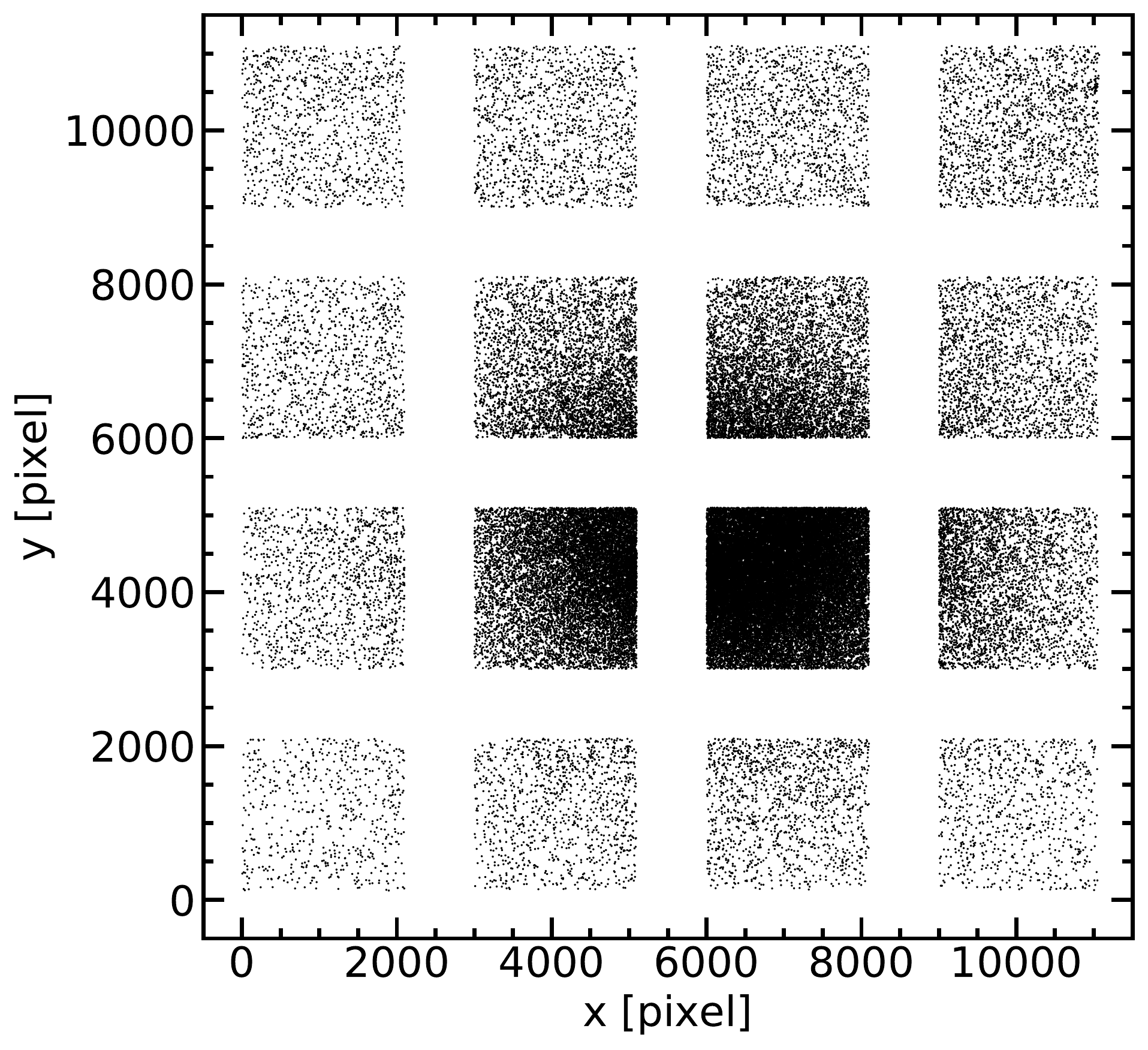} \\
  \includegraphics[width=7.0cm]{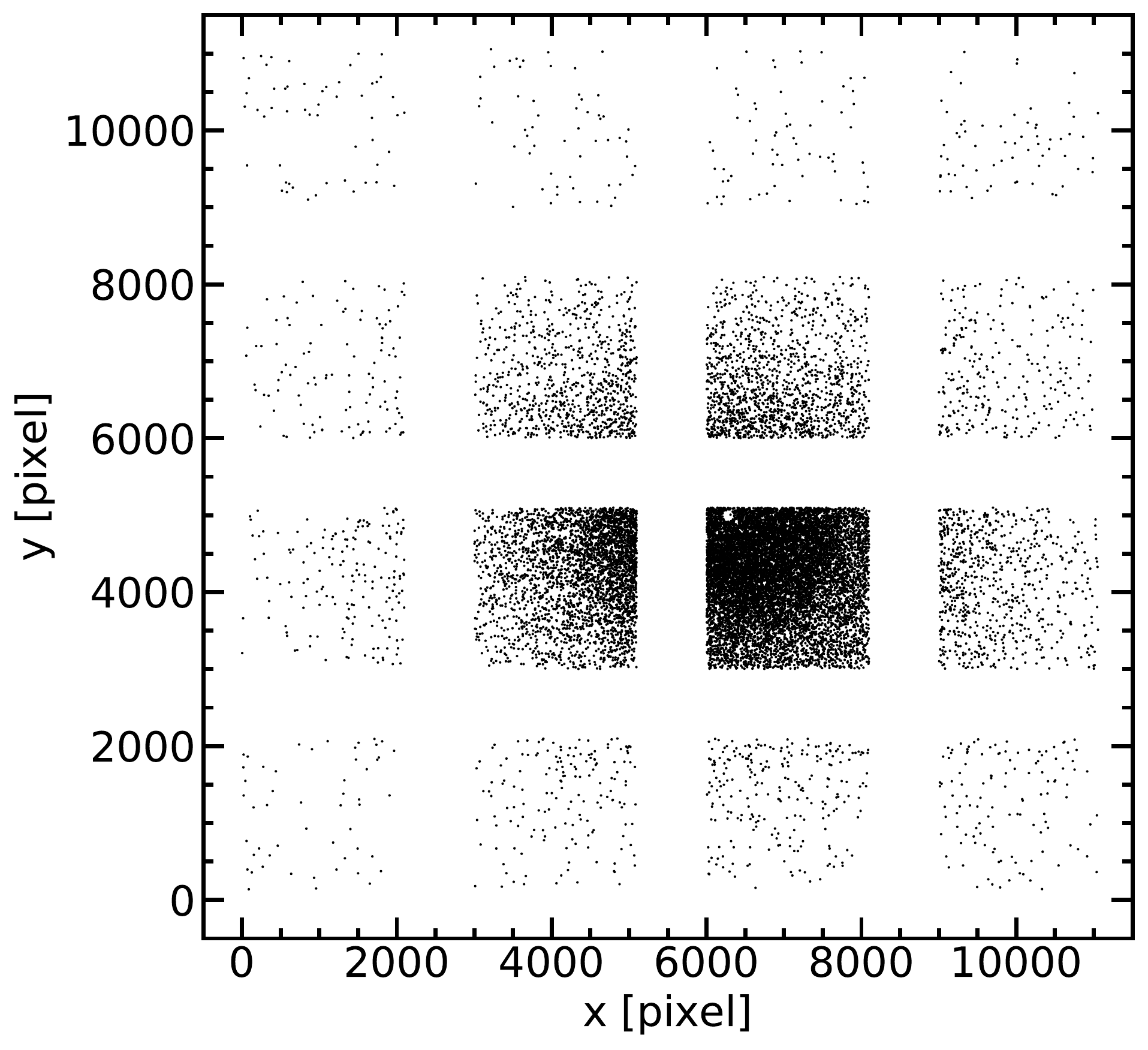} \\
  \includegraphics[width=7.0cm]{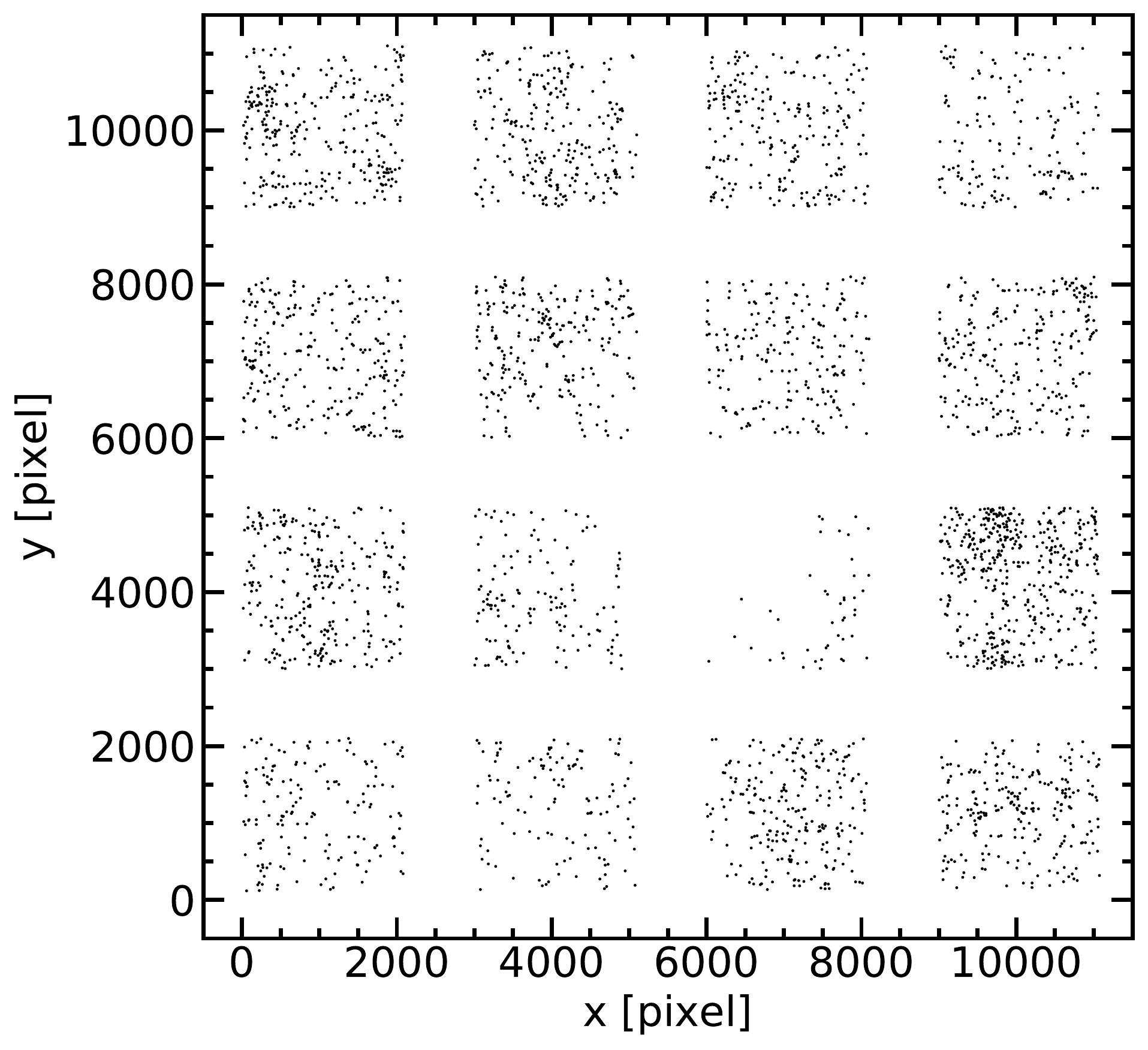} \\
 \end{tabular}
  \caption{\textbf{Top Panel:} Positions on the detectors of all stars detected in the first pawprint observation for our reference epoch TK9. In this image, for demonstration purposes, the detectors are shifted by fixed offsets in the $x$ and $y$ directions and the gaps between them do not reflect the real gaps between the different detectors. For this reason, the cluster appears slightly elongated in this image. In this plot, detector 1 is in the lower left corner and detector 16 is in the upper right corner. The detector numbers increase from left to right and from bottom to top. The $x$ axis goes approximately along the Dec direction and the $y$ axis goes approximately along the RA direction. \textbf{Middle Panel:} As the top panel, but for the cluster stars. \textbf{Bottom Panel:} As the top panel, but for the galaxies.}
   \label{fig:gal_stars}
\end{figure}

As our reference epoch we chose TK9, because it was observed under good conditions (see Table~\ref{tab:multi-epoch}) and it includes the largest number of detected galaxies. In most of the detectors, the number of galaxies is typically between 100 and 400. However, in the detectors that image the innermost parts of 47~Tuc the number of galaxies is reduced significantly, owing to the high crowding in these regions (see Figure~\ref{fig:gal_stars} for the first pawprint exposure of TK9, as an example). There, we only have on the order of 30 galaxies. 
The opposite is true for the distribution of the cluster stars. There are several thousand stars in the detectors that cover the inner parts of the cluster but there are only of the order of 50 stars in the detectors covering the outskirts (see middle panel of Figure~\ref{fig:gal_stars}).

For the matching and transformation, both, galaxy- and star-based, we used the three IRAF tasks \textit{xyxymatch}, \textit{geomap}, and \textit{geoxytran}. The first routine, \textit{xyxymatch}, matches a list of input coordinates with the reference coordinate table. 
The task \textit{geomap} uses the matched coordinate pairs from \textit{xyxymatch} as an input to compute a transformation between the input and reference system. We allowed for a general fit geometry, including a shift, rotation, and scaling, in both the $x$ and $y$ directions. Finally, \textit{geoxytran} transforms the coordinates from the input table to the reference frame using the parameters determined by \textit{geomap}. We performed this transformation procedure for each detector in all pawprints to convert them to the corresponding ones in our reference epoch TK9. Table ~\ref{tab:geomapGalaxies} gives the statistics of the initial transformation using the galaxy frame, resulting from \textit{geomap}. After inspection of the quality of the conversions we found that the $r.m.s.$ residuals in $x$ and $y$ were always below 0.17~pixels, except for epoch TK5, where the $r.m.s.$ was in general above 0.2~pixels. \citet{Cioni16} already noted that there are some problems with this epoch, possibly due to the high airmass. They excluded it from their calculations and we also decided not to use it for further calculations. 

\begin{table} \small
\centering
\caption{Statistics of the results from the $geomap$ transformation using background galaxies for the definition of the common reference frame. The scatter is the standard deviation. \label{tab:geomapGalaxies}} 
\begin{tabular}{@{}l@{ }c@{ }c@{ }c@{ }c@{ }c@{ }c@{ }}
\hline\hline
\noalign{\smallskip}
&\multicolumn{2}{c}{Shift} & \multicolumn{2}{c}{Rotation} & \multicolumn{2}{c}{Magnification}\\
&\multicolumn{2}{c}{(pixel)} & \multicolumn{2}{c}{(degree)} & \multicolumn{2}{c}{}\\
&~~~~$x$~~~~ & ~~$y$~~ & ~~$x$~~ & ~~$y$~~ & ~~$x$~~ & ~~$y$~~\\
\noalign{\smallskip}
\hline
\noalign{\smallskip}
Mean& $-$0.059~~& ~~0.784~~& $-$0.0059~~& ~~0.0012~~&~~0.999996~~&~~1.000023\\
Scatter& ~~0.784~~& ~~1.674~~& ~~0.0045~~& ~~0.0076~~&~~0.000055~~&~~0.000058\\
Maximum& ~~2.929~~& ~~4.504~~& ~~0.0164~~& ~~0.0168~~&~~1.000172~~&~~1.000195\\
\noalign{\smallskip}
\hline
\end{tabular}

\end{table}

We then performed the refined transformation using the catalogs resulting from the initial one as the input and the cluster stars as the reference points. The statistics of this second transformation are listed in Table~\ref{tab:geomapStars}. As can be seen from the table, only small corrections are required. The  $r.m.s.$ residuals in the $x$ and $y$ direction are now below 0.09~pixels.

We repeated the above steps incorporating also quadratic terms in the transformation to explore their effect on the final result. However, we found that allowing for quadratic transformations introduces additional scatter in the proper motion of both the galaxies and the stars. For this reason, we decided to use only the linear transformations.

\begin{table} \small
\centering
\caption{Statistics of the results from the $geomap$ transformation using the stars of the cluster itself for the definition of the common reference frame. The scatter is the standard deviation. \label{tab:geomapStars}}
\begin{tabular}{@{}l@{ }c@{ }c@{ }c@{ }c@{ }c@{ }c@{ }}
\hline\hline
\noalign{\smallskip}
&\multicolumn{2}{c}{Shift} & \multicolumn{2}{c}{Rotation} & \multicolumn{2}{c}{Magnification}\\
&\multicolumn{2}{c}{(pixel)} & \multicolumn{2}{c}{(degree)} & \multicolumn{2}{c}{}\\
&~~~~$x$~~~~ & ~~$y$~~ & ~~$x$~~ & ~~$y$~~ & ~~$x$~~ & ~~$y$~~\\
\noalign{\smallskip}
\hline
\noalign{\smallskip}
Mean& $-$0.001~~& ~~0.013~~& ~~0.0002~~& ~~0.0000~~&~~0.9999997~~&~~1.0000002\\
Scatter& ~~0.040~~& ~~0.037~~& ~~0.0012~~& ~~0.0014~~&~~0.0000242~~&~~0.0000252\\
Maximum& ~~0.168~~& ~~0.179~~& ~~0.0051~~& ~~0.0087~~&~~1.0001103~~&~~1.0001016\\
\noalign{\smallskip}
\hline
\end{tabular}

\end{table}

\subsection{Deriving the Proper Motions}

\begin{figure}
\centering
\includegraphics[width=\columnwidth]{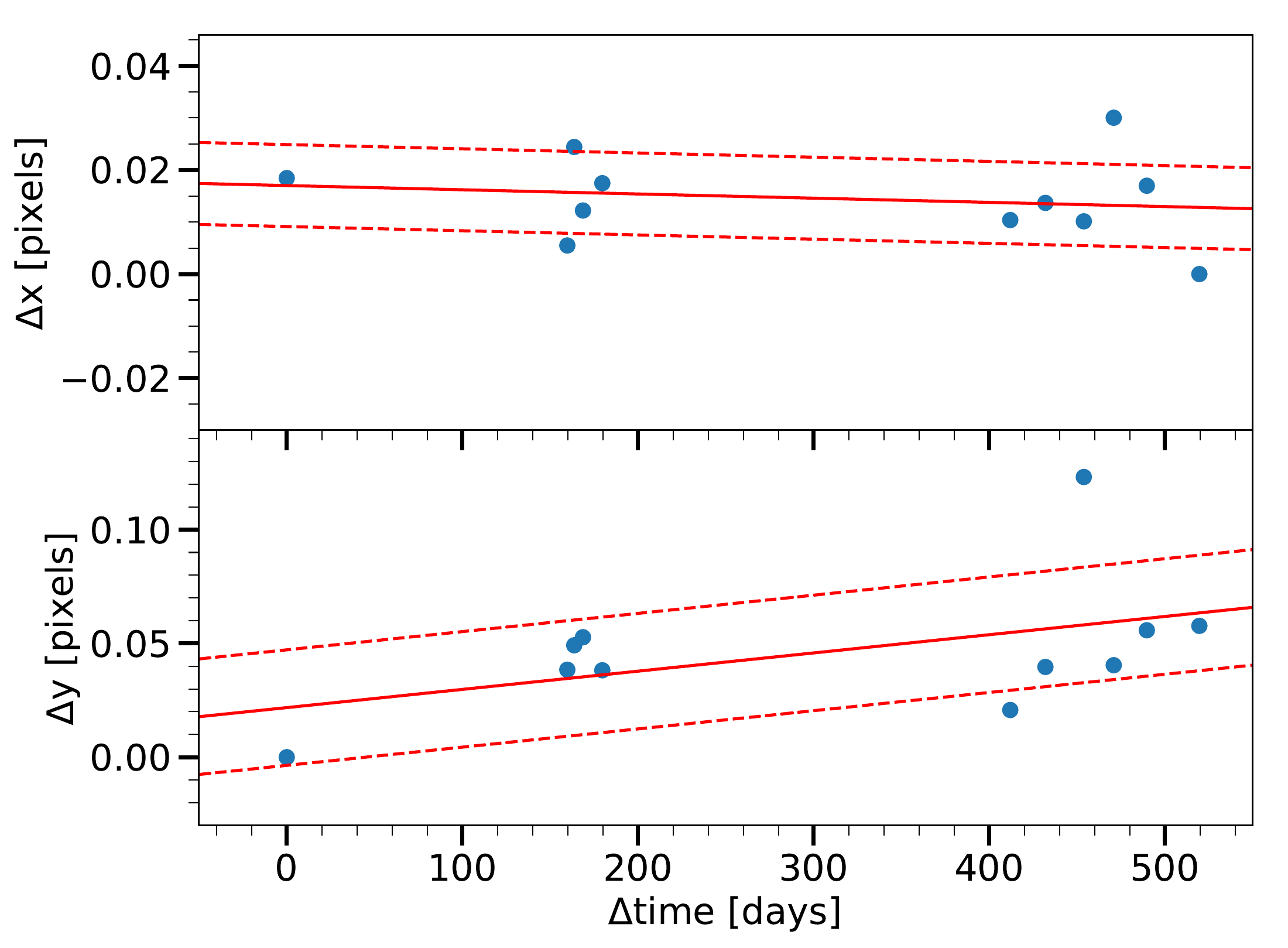}
\caption{Position of a single random star on one detector chip as a function of time (filled circles). The solid line is the best-fitting linear curve to the data. The dashed lines indicate the 1~$\sigma$ distance of the stars from the fitted line. The zero-point of the ordinate is arbitrarily chosen with respect to the minimum value.}
\label{fig:single_star_pm}
\end{figure}

\begin{figure}
\centering
\includegraphics[width=0.9\columnwidth]{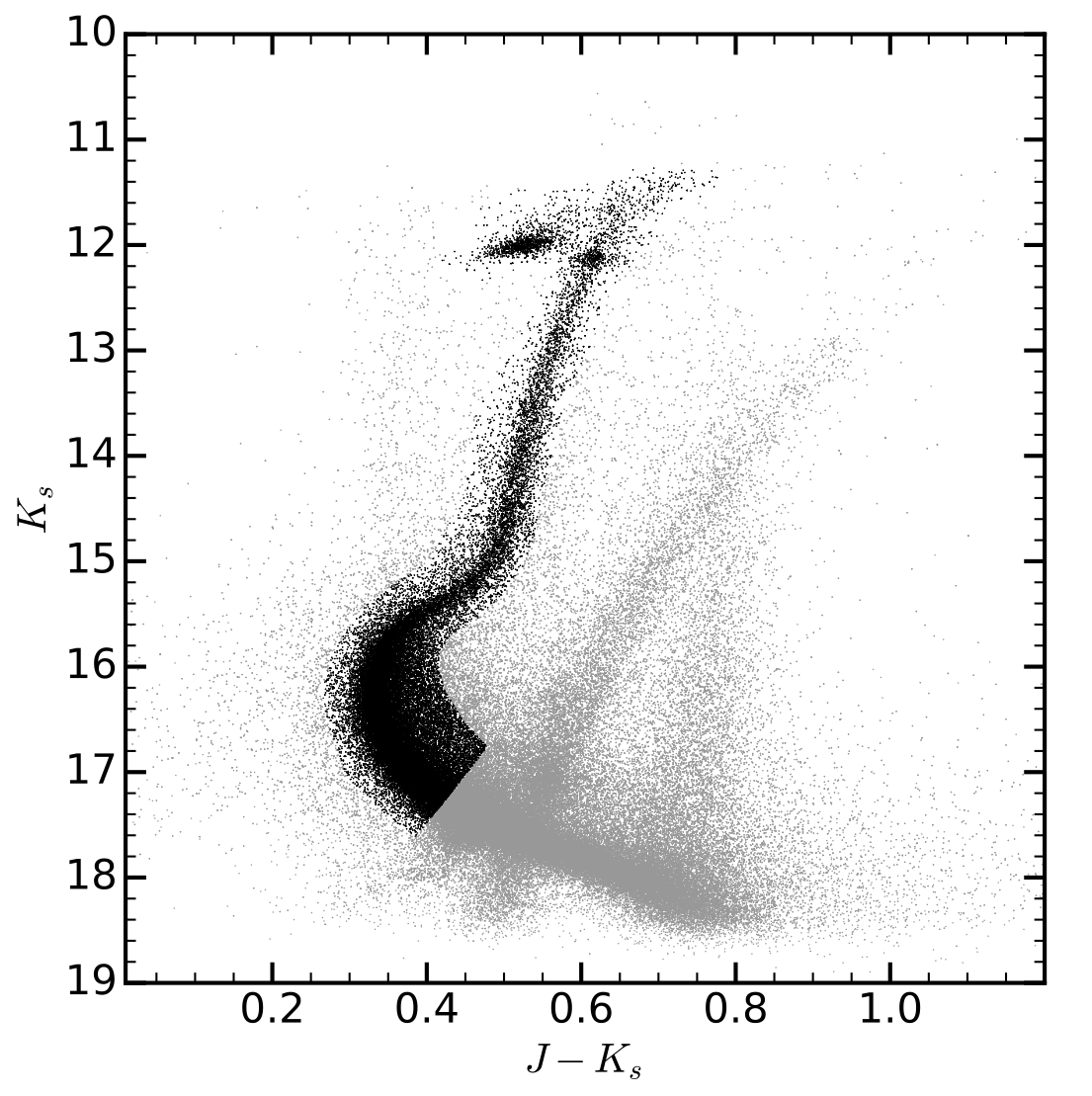}
\caption{CMD of all stars with $K_s$ photometric errors $\sigma(K_s)\leq$0.05~mag. Highlighted as black dots are the stars we selected for our calculation of the proper motion of 47~Tuc.}
\label{fig:mask}
\end{figure}

After the preparatory work described in the previous sections, our final data set for the computation of the proper motions consists of catalogs at 11 different epochs, all shifted to a common frame of reference. We identified for all detectors and pawprints separately all stars that are detected in all epochs (about 60\% of the total number of stars) and created catalogs for each star, containing the various $x$ and $y$ positions on the detector chip and the respective MJDs. Next, we fitted a linear least-squares regression model independently to the $x$ and $y$ values as a function of the MJD. Figure~\ref{fig:single_star_pm} shows, as an example, the $x$ and $y$ measurements of a random star in our sample on a single detector at the various epochs, together with the best-fitting linear model. The slope of the fitted model gives us the proper motions d$x$ and d$y$ of the sources in units of pixels~day$^{-1}$. 

Finally, we transformed the results to a more meaningful physical unit. For this, we first applied the following transformation equations to our values (see also \citealt{Cioni16}) which take into account the pixel scale and the rotation of the detector with respect to the sky coordinates:

\begin{equation}
\mathrm{d}\xi = CD_{1,1}\mathrm{d}x+CD_{1,2}\mathrm{d}y
\end{equation}

\begin{equation}
\mathrm{d}\eta = CD_{2,1}\mathrm{d}x+CD_{2,2}\mathrm{d}y
\end{equation}

We follow the convention introduced by \citet{Cioni16} where d$\xi$ equals $\mu_{\alpha}\mathrm{cos}(\delta)$ and d$\eta$ stands for $\mu_{\delta}$ with $\alpha$ running along the right ascension and $\delta$ along the declination axes.
The four coefficients $CD_{1,1}, CD_{1,2}, CD_{2,1}$ and $CD_{2,2}$ are the transformation matrix elements which account for the rotation of the field on the plane of the sky and for the pixel scale. They can be retrieved from the FITS headers of the detectors of all pawprints of the reference epoch. 
We note that the astrometry of VISTA
pawprints shows a systematic pattern of the order of 10-20 mas coming
from residual World Coordinate System (WCS) errors (see \url{http://casu.ast.cam.ac.uk/surveys-projects/vista/technical/astrometric-properties}).
This effect limits the precision of the proper motion measurements of single objects in this study. But as we will show later (see Section~\ref{sec:compar}) our overall results are in good agreement with recent studies, which suggests that there is no significant systematic offset in the astrometry. We expect the precision of the VISTA astrometric solution to improve in the future since it will be calibrated using Gaia DR2 data.

The above equations transform our proper motion measurements from pixels~day$^{-1}$ to deg~day$^{-1}$, which we finally convert to mas~yr$^{-1}$.
We also applied the same method to our sample of galaxies. We will later use the reflex proper motions of the galaxies as our zero point to determine the motions of the stars.

Our final star catalog of proper motions includes a total of 
277,908 objects. The positions of the six pawprint pointings are designed to ensure the resulting tile automatically observes each object in at least two different pawprints (except in narrow strips at the outside edge of the tiles). A small fraction of objects will be observed more than twice, along edges and corners of detectors. Thus roughly half of the objects in the 6 pawprints making a tile correspond to the same object on the sky, although some objects will appear more than twice. In total, there are 134,605 unique entries in our catalog of stars. The derived proper motions of all stars in this tile are absolute proper motions, since they have been calculated with respect to background galaxies which represent a non-moving reference frame.
We will make the proper motion catalogs publicly available upon request.

\section{Results and Analysis \label{sec:results}}

\subsection{The Proper Motion of 47~Tuc}
\subsubsection{The VMC Survey Data\label{sec:vmc_pm}}

For the calculation of the median proper motion of 47~Tuc 
we applied the same criteria as in Section~\ref{sec:reference} to select the cluster members.
Owing to high crowding in the innermost parts of the cluster, our results suffer from inaccurate measurements of the stellar centroids, even when using PSF photometry. We will therefore exclude the inner 5$\arcmin$ of the cluster from further calculations. 
Furthermore, we restrict our final sample to stars that are within the cluster's tidal radius of $\sim$42$\arcmin$ (we assume the same value as \citealt{Cioni16}). We are left with a total of 78,535 detections of potential cluster members (34,877 unique stars, considering the multiple detections) to calculate the proper motion of the cluster. 
After the star-based transformation, the cluster stars are at rest whereas the galaxies move. The proper motion of the cluster with respect to the galaxies is therefore given by the negative median reflex motion of the background galaxies. 
We  found a median proper motion of 47~Tuc of $(\mu_{\alpha}\mathrm{cos}(\delta),~\mu_{\delta}) = (+5.89, -2.14)$~mas~yr$^{-1}$ with a statistical error of 0.02~mas~yr$^{-1}$ in both directions. These values have been corrected for the median residual motions of the cluster stars in the $\alpha$ and $\delta$ direction (d$\xi = 0.00 \pm \mathbf{0.02}$ and d$\eta = 0.07 \pm \mathbf{0.02}$~mas~yr$^{-1}$) to place them at rest.
The statistical error of the measurements is calculated as the median absolute deviation (MAD) divided by the square root of the numbers of stars. The MAD is defined as the median of the difference between the measurements and its median. Owing to the large numbers of stars used to calculate the absolute proper motion of 47~Tuc, the statistical error is very small. 

In addition to the statistical error, there are also systematic uncertainties which are mostly caused by atmospheric turbulence, the uncertainties in the calibration of the individual detector images using 2MASS stars and the atmospheric differential refraction. To limit the effects of the last mentioned contribution, we used for our calculations only observations in the $K_s$ band taken at similar air masses (the epoch with the largest air mass was removed due to large residuals in the transformation). For a detailed discussion of the various sources of errors see \citet{Cioni16}. 
We can estimate the systematics from the uncertainties in the cluster stars based reference frame. Looking at the proper motions of the stars as a function of detector number (see Figure~\ref{fig:pm_vs_detector}), we found a scatter of the median motion of 0.50~mas~yr$^{-1}$ in the RA direction and 0.31~mas~yr$^{-1}$ in the Dec direction, resulting in a larger uncertainty in the first direction. A good measure of the statistical uncertainties of the proper motions results would be the scatter of the median proper motions of the stars scaled to the square root of the number of the detectors. This results in an error of 0.13~mas~yr$^{-1}$ in the RA direction and 0.11~mas~yr$^{-1}$ in the Dec direction, taking also into account the residual motion of the stars in the Dec direction. This demonstrates that the precision of the measurements is limited by the systematic uncertainties. Our final result of the absolute proper motion of 47~Tuc is $(\mu_{\alpha}\mathrm{cos}(\delta),~\mu_{\delta}) = (+5.89 \pm 0.02~\mathrm{(statistical)} \pm 0.13~\mathrm{(systematic)}, -2.14 \pm 0.02~\mathrm{(statistical)} \pm 0.08~\mathrm{(systematic)})$~mas~yr$^{-1}$.
VISTA’s optical design results in a radially symmetric cubic (pincushion) scale distortion at the VIRCAM detectors which results in a
smaller pixel size at the edges of a pawprint \citep[see][]{Sutherland15}. The maximum difference in pixel scale between the central parts and the outermost parts in the pawprint is of the order of 2 per cent which translates in a difference in proper motion of 2 per cent, as well. Since the mean variation in pixel size is much smaller and we averaged for our analysis the results from many stars spread over a large area within the pawprint, we can neglect here the effect of the distortion.

We also looked at the results of the 47~Tuc's proper motion when only the galaxy-based transformation to a common frame of reference is used. In this case, the galaxies are at rest and the stars of the clusters move and for the median proper motion of 47~Tuc we found $(\mu_{\alpha}\mathrm{cos}(\delta),~\mu_{\delta}) = (+5.44 \pm 0.02~\mathrm{(statistical)} \pm 0.15~\mathrm{(systematic)}, -2.52 \pm 0.02~\mathrm{(statistical)} \pm 0.28~\mathrm{(systematic)})$~mas~yr$^{-1}$. To put the galaxies at rest, these values have been corrected for the median residual reflex motion of the galaxies in the $\alpha$ and $\delta$ direction (d$\xi = -0.12 \pm 0.12$ and d$\eta = 0.01 \pm 0.12$~mas~yr$^{-1}$).
When we compare the galaxy-based measurement of 47~Tuc's proper motion with the star-based value, we can see that both components of the former are smaller. 
In the Dec direction, the difference is somewhat larger than the 1$\sigma$ uncertainty, whereas the two values differ by about $2.3\sigma$ in the RA direction. The number of reference objects to calculate the relative proper motion is significantly different in the two methods. 
This allows for a more well-defined reference system in the star-based method and results in a final value 
that differs moderately from the galaxy-based value.
But as we will see later in Section~\ref{sec:compar}, the values from both methods are compatible with the range of results from recent studies.

\begin{figure}
\centering
\includegraphics[width=0.95\columnwidth]{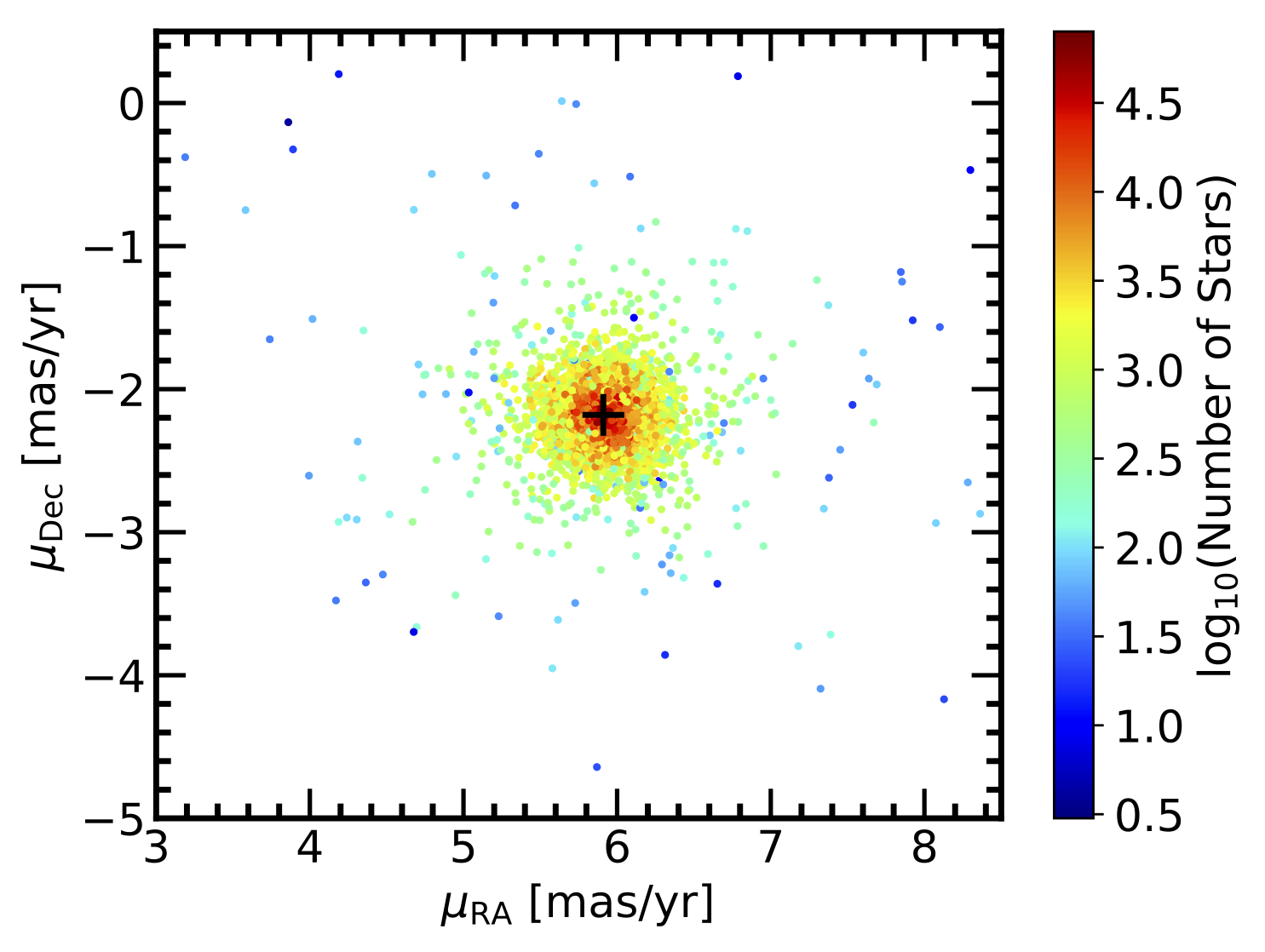}
\caption{Resulting proper motion values of 47~Tuc when choosing a randomly selected sub-sample of cluster stars. Each point represents an individual result, and is color-coded by the number of stars used for the calculation. The 'real' value using all stars is indicated by the black cross.}
\label{fig:pm_vs_number}
\end{figure}

\begin{figure}
\centering
\includegraphics[width=1.0\columnwidth]{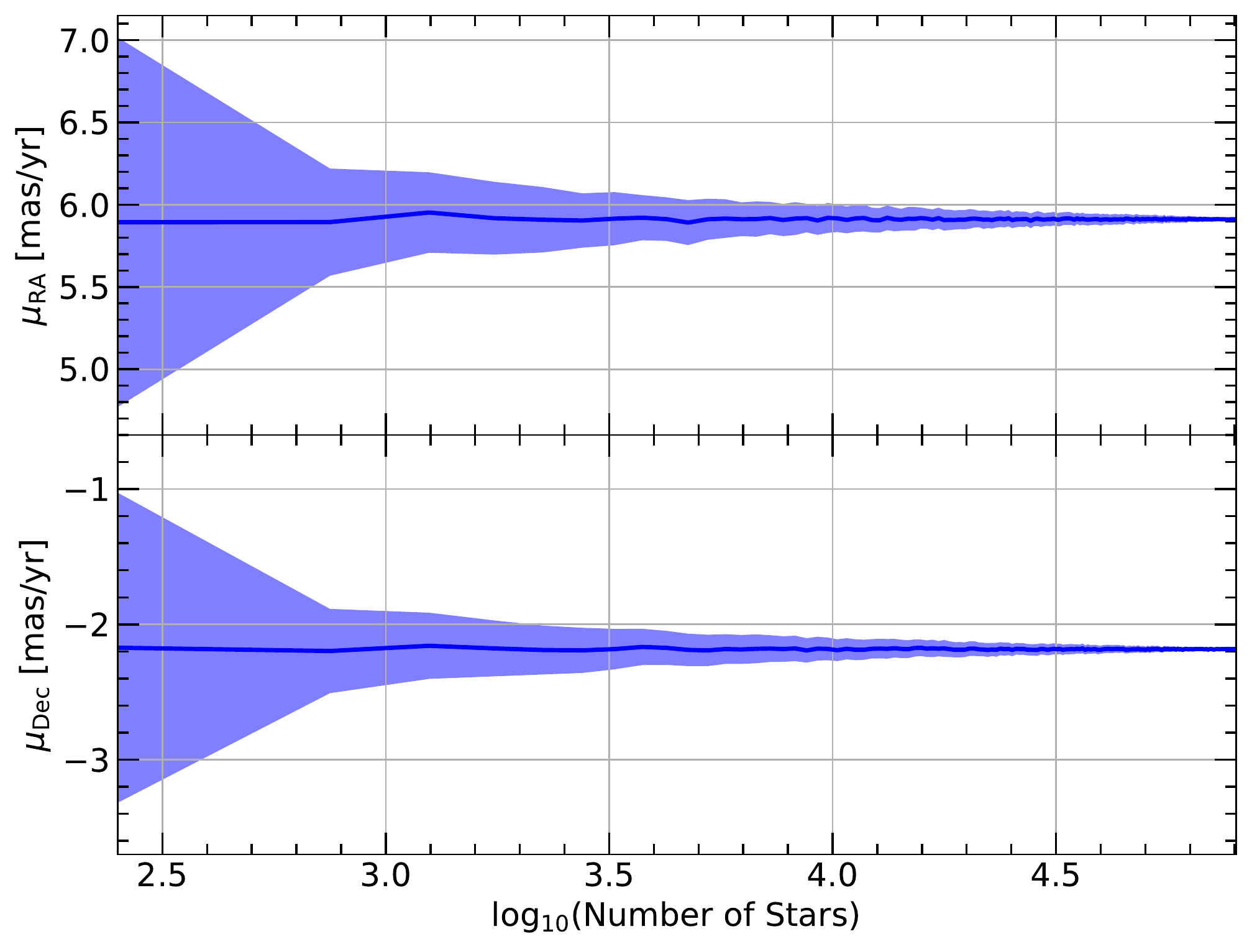}
\caption{Median value of the sub-sample results shown in Figure~\ref{fig:pm_vs_number} in bins of 500 stars (solid blue line), along with its standard deviation (blue shaded area). The bottom panel is for the proper motion in the Dec direction and the top panel for the proper motion in the RA direction.}
\label{fig:median_vs_number}
\end{figure}

Next, we explore how stable our result is with respect to the size of the sample of stars used. This will help us to assess the reliability of our measurements in regions where fewer stars are available, e.g. the extra-tidal region of the cluster (see above). For this, we created 70,000 (approximate number of stars in our original sample) sub-samples of stars from our original cluster catalog. Each consisted of a random number of stars which were drawn randomly from the original list of cluster stars. We then calculated the median proper motion of every sub-sample. The results are shown in Figure~\ref{fig:pm_vs_number}. Each point represents an individual measurement and is color-coded by the number of stars used. The black cross marks the proper motion value that is derived from the entire list of cluster stars. As expected, the fewer stars are used for the calculation, the less accurate the result gets. Especially, for sample sizes of less than a few hundred stars, the original signal imprinted in the data gets lost in the measurement errors of the individual stars, and the final result becomes more or less random. To quantify the reliability of the results using variable numbers of stars, we calculated the medians and standard deviations of all values in bins of 500 stars (see Figure~\ref{fig:median_vs_number}). For a sample size of 2000 stars the 1 $\sigma$ scatter is of the order of 0.3 $-$ 0.4~mas~yr$^{-1}$ around the original value. This scatter decreases to about 0.1~mas~yr$^{-1}$ for catalogs of 10,000 stars. From this test, we conclude that for a reasonable and reliable result, we need sample sizes of, at least, 1500 $-$ 2000 stars.

Finally, we examined the reflex proper motions of the background galaxies and the proper motions of the cluster stars as a function both of the $J-K_s$ color and the distance to the center of 47~Tuc to look for any systematic trends in the motion of these objects. The corresponding figures are presented in the Appendix (Figure~\ref{fig:pm_vs_colour} and \ref{fig:pm_vs_radius}) and show the distributions of the individual proper motions together with the running mean. 
In almost all cases, we do not find any significant trend of the proper motions as a function of the color or distance from the cluster center. The only exception where a trend might be present is the Dec component of the galaxies' reflex motion as a function of the distance from the cluster center (see top right panel of Figure~\ref{fig:pm_vs_radius}). There, the mean reflex motion is systematically below zero at radii smaller than $\sim$33$\arcmin$ and above zero at larger radii. A Spearman correlation test reveals that in this case, the correlation coefficient is about 0.02 whereas in the other cases, it is always well below 0.01. Since we are only interested in the overall movement of 47~Tuc and the median reflex motion of the galaxies across all detectors is close to zero (d$\eta = 0.01 \pm 0.12$~mas~yr$^{-1}$), this slight trend has no significant effect on our final results.

\subsubsection{The UCAC 5 Catalog}

\begin{figure}
 \begin{tabular}{c}
  \includegraphics[width=7.0cm]{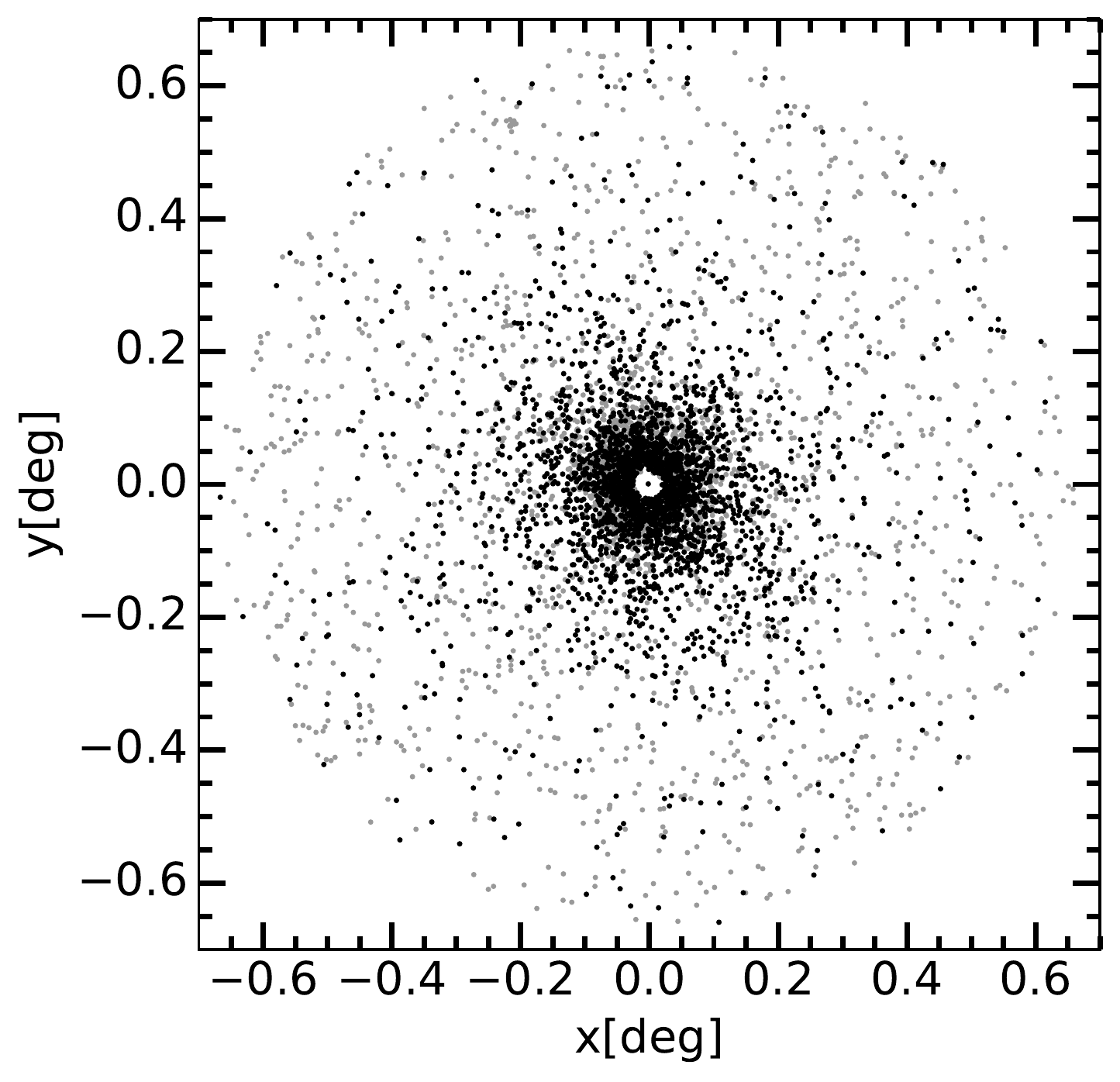} \\
  \includegraphics[width=7.0cm]{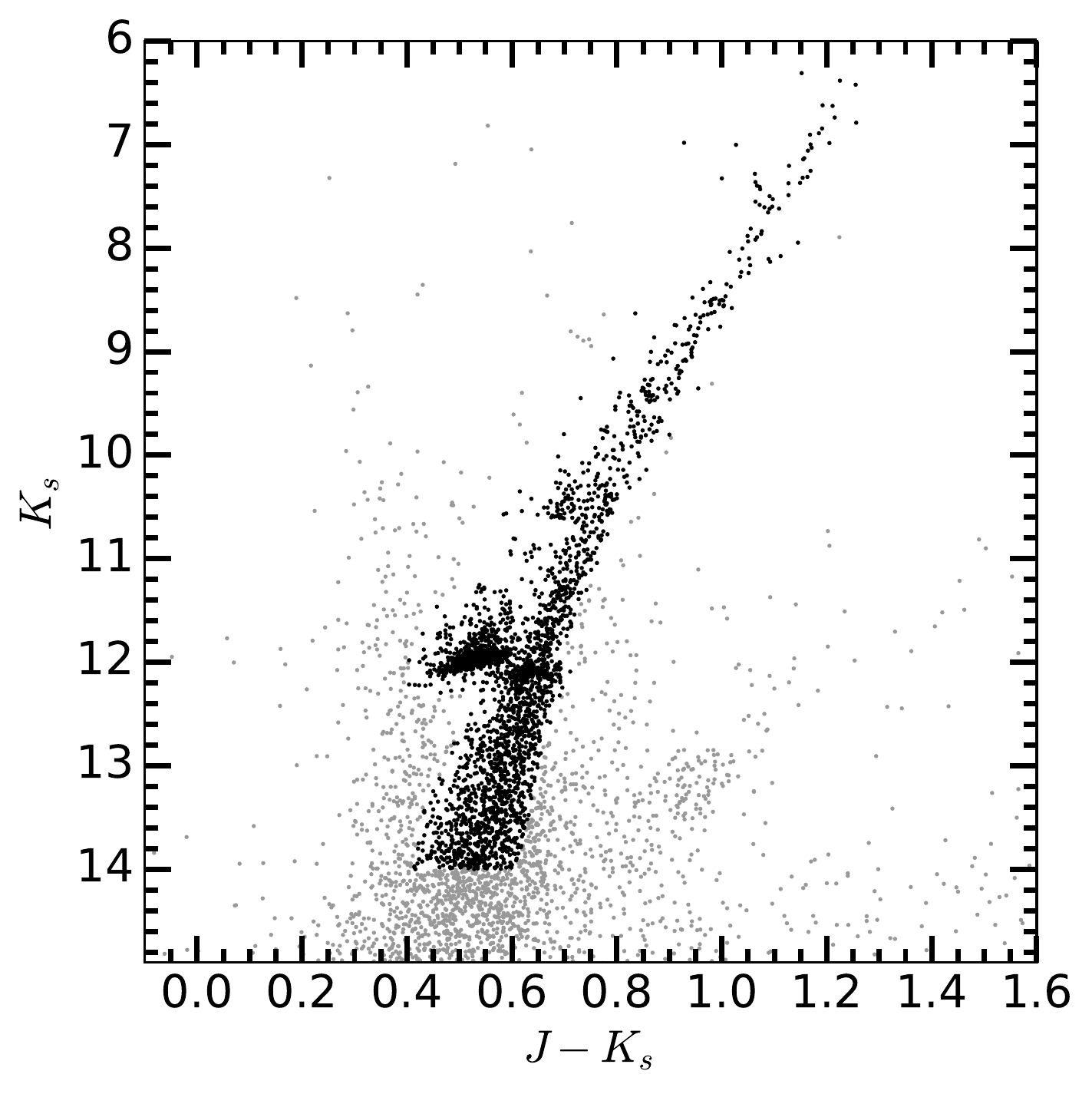} \\
 \end{tabular}
  \caption{\textbf{Top Panel:}  Spatial distribution of stars in the UCAC5 catalog within 42$\arcmin$ of the center of 47~Tuc. Stars selected for the proper motion calculation are indicated by black dots. \textbf{Bottom Panel:} $K_s$ vs $J-K_s$ CMD of the stars shown in the top panel.}
   \label{fig:ucac5}
\end{figure}

In addition to the VMC result, we also used data from the newest US Naval Observatory CCD Astrograph Catalog (UCAC5) proper motion catalog which also includes data from \textit{Gaia} DR1 \citep{Zacharias17} for an independent measurement of 47~Tuc's proper motion. From the catalog\footnote{\url{http://dc.zah.uni-heidelberg.de/ucac5/q/cone/form}}, we queried all stars that reside within 42$\arcmin$ of the cluster center. Similarly to the VMC data, we masked the main CMD features of the cluster to get a cleaner sample of cluster stars. We selected stars that follow the RGB, RC and asymptotic giant branch of 47~Tuc. Additionally, we limited our sample to stars with magnitudes brighter than 14.0 in the $K_s$ band, which is about one magnitude above the sensitivity limit of the data. Our final list contains about 2750 stars. The top panel of Figure~\ref{fig:ucac5} shows the spatial distribution of the queried stars. The black dots indicate the stars we selected for our proper motion calculation. The bottom panel of Figure~\ref{fig:ucac5} shows the same stars in a CMD using the 2MASS $J$ and $K_s$ filters. Again, the selected stars are denoted by black dots. Using this final list of stars, we found a median proper motion of the stars in 47~Tuc of $(\mu_{\alpha}\mathrm{cos}(\delta),~\mu_{\delta}) = (+5.30, -2.70)$~mas~yr$^{-1}$ with a statistical error of 0.03~mas~yr$^{-1}$ in both directions. 
The precision, in terms of the statistical uncertainties, of this measurement is comparable to the one from the VMC data, although we calculated the proper motion from a much smaller number of stars using the UCAC5 catalog. The reason for this is that the proper motions in the UCAC5 catalog are calculated using a much longer time baseline ($\sim$14 years). However, \citet{Zacharias17} noted that the proper motions suffer from systematic uncertainties which can be as large as 0.7~mas~yr$^{-1}$. The final result of the proper motion of 47~Tuc from the UCAC5 data catalog is $(\mu_{\alpha}\mathrm{cos}(\delta),~\mu_{\delta}) = (+5.30 \pm 0.03~\mathrm{(statistical)} \pm 0.70~\mathrm{(systematic)}, -2.70 \pm 0.03~\mathrm{(statistical)} \pm 0.70~\mathrm{(systematic)})$~mas~yr$^{-1}$, which agrees with the value from the VMC data within the uncertainties.
The UCAC5 catalog would be only suitable for limited dynamical studies of the Magellanic Clouds. Since its sensitivity limit is at $K_s\sim$15~mag (the RC of the SMC is more than 2 magnitudes fainter, see Figure~\ref{fig:hess}) it reaches only the brightest evolved stars (upper RGB, AGB) in the Clouds.

\subsection{Comparison with Literature Values}
\subsubsection{Results from Aperture Photometry}

\citet{Cioni16} already calculated the proper motion of the different stellar populations present in VMC tile SMC 5\_2, including 47~Tuc, using aperture photometry measurements. For the cluster, they found $(\mu_{\alpha}\mathrm{cos}(\delta),~\mu_{\delta}) = (+7.26 \pm 0.03~\mathrm{statistical} \pm 0.18~\mathrm{(systematic)}, -1.25 \pm 0.03~\mathrm{(statistical)} \pm 0.18~\mathrm{(systematic)})$~mas~yr$^{-1}$. 
This is incompatible with the values found in this work using PSF photometry. When using the stellar based transformation, the difference in $\mu_{\alpha}\mathrm{cos}(\delta)$ is 6.2$\sigma$ and in $\mu_{\delta}$ it is 4.5$\sigma$ . When using the galaxies as the frame of reference, the results differ by 8$\sigma$ in $\mu_{\alpha}\mathrm{cos}(\delta)$ and in $\mu_{\delta}$ by 4$\sigma$. In the following we will discuss several reasons which might be responsible for this discrepancy. We will base our discussion here on our result from the galaxy based transformation, since \citet{Cioni16} also use the galaxies as their reference points.

As a first point, the different time baseline used to calculate the proper motions is to be mentioned. \citet{Cioni16} only used the deep exposures which span in total about 12 months. In our study, we also include the first shallow epoch which extends the time baseline to about 17 months. To assess the effect of the longer time interval, we re-calculated the proper motions using only the 10 deep observations which yields $(\mu_{\alpha}\mathrm{cos}(\delta),~\mu_{\delta}) = (+5.43, -1.41)$~mas~yr$^{-1}$. The shorter time baseline shifts our result closer to the one obtained by \citet{Cioni16} in the Dec direction but there is only a slight change in the RA direction. 

Second, the selection of the stars used to calculate the absolute motion of 47~Tuc is different in both studies. For their calculations, \citet{Cioni16} used all stars between 10$\arcmin$ and 60$\arcmin$ from the center of the cluster that are within certain CMD regions (see their Figure~6) which results in approximately 70,000 measurements in total. To compare the effect of the different stellar samples, we applied the same selection criteria to both proper motion catalogs. When we applied the same radial range and CMD region masks as \citet{Cioni16} to our data using our original set of epochs, we found $(\mu_{\alpha}\mathrm{cos}(\delta),~\mu_{\delta}) = (+4.64, -2.61)$~mas~yr$^{-1}$. With the shorter time baseline, the proper motion now becomes  $(\mu_{\alpha}\mathrm{cos}(\delta),~\mu_{\delta}) = (+4.38, -1.50)$~mas~yr$^{-1}$. This value is now smaller than our original result, most likely because the new sample also includes RC stars from the SMC which have smaller proper motions.
We also applied our selection of 47~Tuc stars to the \citet{Cioni16} catalog, which gives $(\mu_{\alpha}\mathrm{cos}(\delta),~\mu_{\delta}) = (+7.06, -1.36)$~mas~yr$^{-1}$. 
The various selections of stars change the overall result but cannot account for the entire difference of the outcomes of the two methods which suggests an intrinsic difference in the two catalogs.

The third reason is the different determination of the stellar centroids in the standard VDFS pipeline and the PSF fitting technique which becomes especially evident in regions with a high stellar density, like in the inner part of the cluster. We cross-correlated single-epoch PSF and aperture photometry stellar catalogs of various detectors, to quantify the differences in stellar positions resulting from the two methods. We found that, regardless of stellar crowding, there are systematic shifts both in the $x$ and $y$ directions of the stellar positions, within a single detector. These shifts are different at different epochs and usually vary between 0~pixels and $(1-2)\times10^{-2}$~pixels but there are also single cases where the shifts are as large as $\sim5\times10^{-2}$~pixels. 
The differences in stellar positions in both directions have a 1$\sigma$ scatter which is about 0.2~pixels in crowded regions and $\sim$0.06~pixels in less crowded regions. We also compared the proper motion results from both methods in regions with a low stellar density. For this, we selected stars with distances larger than 40$\arcmin$ from the center of 47~Tuc and chose the same CMD region masks as \citet{Cioni16} to select SMC stars along the RGB and RC. Again, using only the 10 deep observations, we found a proper motion of the SMC stars of $(\mu_{\alpha}\mathrm{cos}(\delta),~\mu_{\delta})= (+1.23, -0.61)$~mas~yr$^{-1}$ with a statistical uncertainty of 0.07~mas~yr$^{-1}$. This is similar to the results by \citet{Cioni16} who found  $(\mu_{\alpha}\mathrm{cos}(\delta),~\mu_{\delta}) = (+1.16, -0.81)$~mas~yr$^{-1}$ which suggest that both methods work similarly well in less crowed regions.

Finally, \citet{Cioni16} used a different set of galaxies for the alignment of the observations at different epochs. We have chosen our sample of galaxies applying stricter selection criteria and have a smaller galaxy sample size. This can result in a different transformation of the various observations. As a consequence the residual motions of the galaxies, which are subtracted from the proper motions of the stars, are not the same. The resulting reflex proper motion of the galaxies in \citet{Cioni16} was $(\mu_{\alpha}\mathrm{cos}(\delta),~\mu_{\delta}) = (-0.45 \pm 0.12, -0.15 \pm 0.12)$~mas~yr$^{-1}$, whereas in this work it was $(\mu_{\alpha}\mathrm{cos}(\delta),~\mu_{\delta}) = (-0.12 \pm 0.12, 0.01 \pm 0.12)$~mas~yr$^{-1}$. 

\subsubsection{Other Results from the Literature \label{sec:compar}}

In this section we compare our results with other proper motion measurements from the literature. The most recent one is presented by \citet{Narloch17} who used data from the 1-m Swope telescope of Las Campanas observatory. They calculated the mean proper motion of 47~Tuc with respect to the motion of the SMC using two different fields with time baselines of $\sim$11 and $\sim$17 years, respectively. Adopting a mean value for the motion of the SMC from various measurements in the literature, the authors found for the absolute proper motion of the cluster $(\mu_{\alpha}\mathrm{cos}(\delta),~\mu_{\delta}) = (+5.376 \pm 0.032, -2.216 \pm 0.028)$~mas~yr$^{-1}$. 
Their value agrees with our measurement within the one $\sigma$ uncertainties in the Dec direction but is larger by about 3.8$\sigma$ in the RA direction.

\citet{Watkins17} used data from the TGAS catalog to study the motion and distances of a sample of five globular clusters. They calculated the weighted mean of five identified member stars of 47~Tuc from the TGAS catalog and found as a result $(\mu_{\alpha}\mathrm{cos}(\delta),~\mu_{\delta}) = (+5.50 \pm 0.70, -3.99 \pm 0.55)$~mas~yr$^{-1}$. The authors state that the uncertainties in their measurements are dominated by random errors and the systematic errors are much smaller than those. Our result for the cluster motion is consistent with the finding of \citet{Watkins17} in the RA direction within the errors. However, in the Dec direction the results disagree by about 3.3$\sigma$. This discrepancy might be due to selection effects since the measurement in \citet{Watkins17} results from only five cluster stars.

\citet{Freire03} and, in a follow-up study, \citet{Freire17} observed millisecond pulsars within 47~Tuc. From a weighted average of the proper motions of these pulsars the authors determined the motion of the cluster in their first paper as  $(\mu_{\alpha}\mathrm{cos}(\delta),~\mu_{\delta}) = (5.3 \pm 0.6, -3.3 \pm 0.6)$~mas~yr$^{-1}$. In the second paper, where the long-term observations of the pulsars are presented, \citet{Freire17} found $(\mu_{\alpha}\mathrm{cos}(\delta),~\mu_{\delta}) = (5.00 \pm 0.14, -2.84 \pm 0.12)$~mas~yr$^{-1}$ for the proper motion of the cluster. 
In the RA direction, the older measurement is in agreement with our result whereas the updated value found by \citet{Freire17} is about 4.6$\sigma$ away. In the Dec direction, both values from Freire et al. are smaller than ours. They differ by about 1.9$\sigma$ \citep{Freire03} and 4.9$\sigma$ \citep{Freire17}.

\citet{AndersonKing03} used observations taken with the \textit{Hubble Space Telescope} (HST) to derive the proper motion of 47~Tuc. They used two different fields across the cluster, each of which was observed twice, with a time baseline of 5 and 6 years, respectively. As a reference frame they used the background stars of the SMC, so the resulting proper motion of the cluster depends on the assumed value for the motion of the SMC. \citet{AndersonKing03} found a motion of the SMC with respect to 47~Tuc of $(\mu_{\alpha}\mathrm{cos}(\delta),~\mu_{\delta}) = (-4.716 \pm 0.035, +1.357 \pm 0.021)$~mas~yr$^{-1}$. For the motion of the SMC they took the results from \citet{Irwin99} who give $(\mu_{\alpha}\mathrm{cos}(\delta),~\mu_{\delta}) = (0.92 \pm 0.20, -0.69 \pm 0.20)$~mas~yr$^{-1}$. This results in an absolute proper motion of 47~Tuc of $(\mu_{\alpha}\mathrm{cos}(\delta),~\mu_{\delta}) = (5.64 \pm 0.20, -2.05 \pm 0.20)$~mas~yr$^{-1}$. Combining the results of \citet{AndersonKing03} with more recent measurements of the motion of the SMC from \citet{Kallivayalil06} or \citet{Cioni16} gives $(\mu_{\alpha}\mathrm{cos}(\delta),~\mu_{\delta}) = (5.88 \pm 0.20, -2.53 \pm 0.20)$~mas~yr$^{-1}$ \citep{Kallivayalil06} and $(\mu_{\alpha}\mathrm{cos}(\delta),~\mu_{\delta}) = (5.88 \pm 0.20, -2.17 \pm 0.20)$~mas~yr$^{-1}$ \citep{Cioni16}, respectively. 
Using the updated values for the SMC motion, the result in the RA direction from \citet{AndersonKing03} is in very good agreement with our measurement. Also in the Dec direction both values are consistent using the SMC motion from \citet{Cioni16} but they differ by 1.9 $\sigma$ assuming the value from \citet{Kallivayalil06}.

\citet{Odenkirchen97} used data from the \textit{Hipparcos} mission to measure the proper motions of 15 globular clusters, including 47~Tuc for which they found  $(\mu_{\alpha}\mathrm{cos}(\delta),~\mu_{\delta}) = (7.0 \pm 1.0, -5.3 \pm 1.0)$~mas~yr$^{-1}$. They estimated the error in the absolute proper motion from the dispersion of the proper motions of cluster stars and the accuracy of the \textit{Hipparcos} reference frame. 
In the RA direction, the value obtained by \citet{Odenkirchen97} is larger than ours and just outside the 1$\sigma$ uncertainty. In the Dec direction, our measurement is larger by about $3.2\sigma$.

\citet{CudworthHanson93} found a proper motion of 47~Tuc of $(\mu_{\alpha}\mathrm{cos}(\delta),~\mu_{\delta}) = (3.4 \pm 1.7, -1.9 \pm 1.5)$~mas~yr$^{-1}$ when converting the relative motion of the cluster with respect to background stars to an absolute proper motion where the given error is the combined error of the cluster's relative proper motion uncertainty and the error of the proper motion of the reference stars. 
Our result agrees with the one found by \citet{CudworthHanson93} within the 1$\sigma$ uncertainty in the Dec direction and is larger by about 1.5$\sigma$ in the RA direction.

\begin{table*} \small
\centering
\caption{Proper Motion measurements of 47~Tuc from the literature.  \label{tab:47tucPM}}
\begin{tabular}{c c c} 
\hline\hline
\noalign{\smallskip}
$\mu_{\alpha}\mathrm{cos}(\delta)$ & $\mu_{\delta}$ & Reference\\
(mas~yr$^{-1}$) & (mas~yr$^{-1}$) & \\
\noalign{\smallskip}
\hline
\noalign{\smallskip}

+5.89 $\pm$ 0.13 & $-$2.14 $\pm$ 0.11 & This work (VMC data)\tablefootmark{a}\\
+5.44 $\pm$ 0.15 & $-$2.52 $\pm$  0.28 & This work (VMC data)\tablefootmark{b}\\
+5.30 $\pm$ 0.70 & $-$2.70 $\pm$  0.70 & This work (UCAC5 data)\\
+5.376 $\pm$ 0.032 & $-$2.216 $\pm$ 0.028 & \citet{Narloch17}\\ 
+5.00 $\pm$ 0.14 & $-$2.84 $\pm$ 0.12 & \citet{Freire17}\\
+5.50 $\pm$ 0.70 & $-$3.99 $\pm$ 0.55 & \citet{Watkins17}\\
+7.26 $\pm$0.18 & $-$1.25 $\pm$ 0.18 & \citet{Cioni16}\\
+5.88 $\pm$ 0.20 & $-$2.53 $\pm$ 0.20 & \citet{AndersonKing03}\\
                             &                                 & Using the SMC motion from \citet{Kallivayalil06}\\
+5.88 $\pm$ 0.20 & $-$2.17 $\pm$ 0.20 & \citet{AndersonKing03}\\
                             &                                  & Using the SMC motion from \citet{Cioni16}\\
+5.3 $\pm$ 0.6 & $-$3.3 $\pm$ 0.6 & \citet{Freire03}\\
+7.0 $\pm$ 1.0 & $-$5.3 $\pm$ 1.0 & \citet{Odenkirchen97}\\
+3.4 $\pm$ 1.7 & $-$1.9 $\pm$ 1.5 & \citet{CudworthHanson93}\\

\noalign{\smallskip}
\hline
\end{tabular}
\tablefoot{
\tablefoottext{a}{Common reference frame defined by the cluster stars}\\
\tablefoottext{b}{Common reference frame defined by background galaxies}\\

}
\end{table*}

\begin{figure}
\centering
\includegraphics[width=0.95\columnwidth]{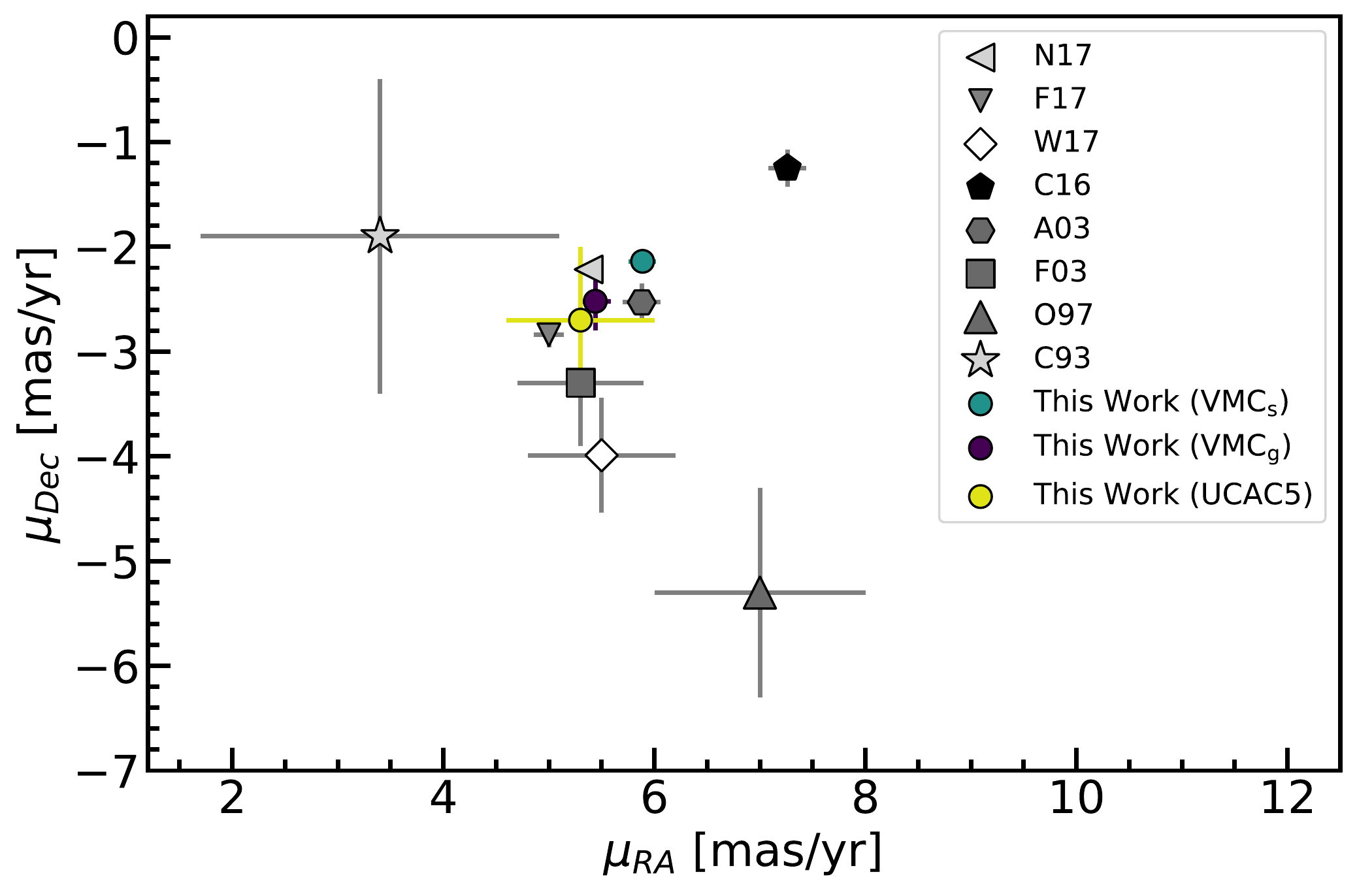}
\caption{Comparison of the median proper motions presented in this paper with values from the literature. Where no uncertainty is shown it is smaller than the size of the symbol. The abbreviations in the legend are the following: N17: \citet{Narloch17}; F17: \citet{Freire17}; W17: \citet{Watkins17}; C16: \citet{Cioni16}; A03: \citet{AndersonKing03}, using the SMC motion from \citet{Kallivayalil06}; F03: \citet{Freire03}; O97: \citet{Odenkirchen97}; C93: \citet{CudworthHanson93}.VMC$_{\mathrm{g}}$: from the VMC data using a galaxy-based reference frame for the transformations between singe epochs; VMC$_{\mathrm{s}}$: from the VMC data using a star-based reference frame for the transformations between singe epochs; UCAC5: from the UCAC5 catalog.}
\label{fig:pm_compar}
\end{figure}

The different measurements, together with the results obtained in this paper, are summarized in Table~\ref{tab:47tucPM} and also illustrated in Figure~\ref{fig:pm_compar}. 
We can see that our measurement is closest to the results obtained by \citet{Narloch17} and \citet{AndersonKing03}. Additionally, our values for the motion in the RA direction agree very well with the ones obtained by \citet{Freire03} and \citet{Watkins17}, although our value in the Dec direction is larger then theirs.

\subsection{The Orbit of 47 Tuc}

\begin{figure}
\centering
\includegraphics[width=0.95\columnwidth]{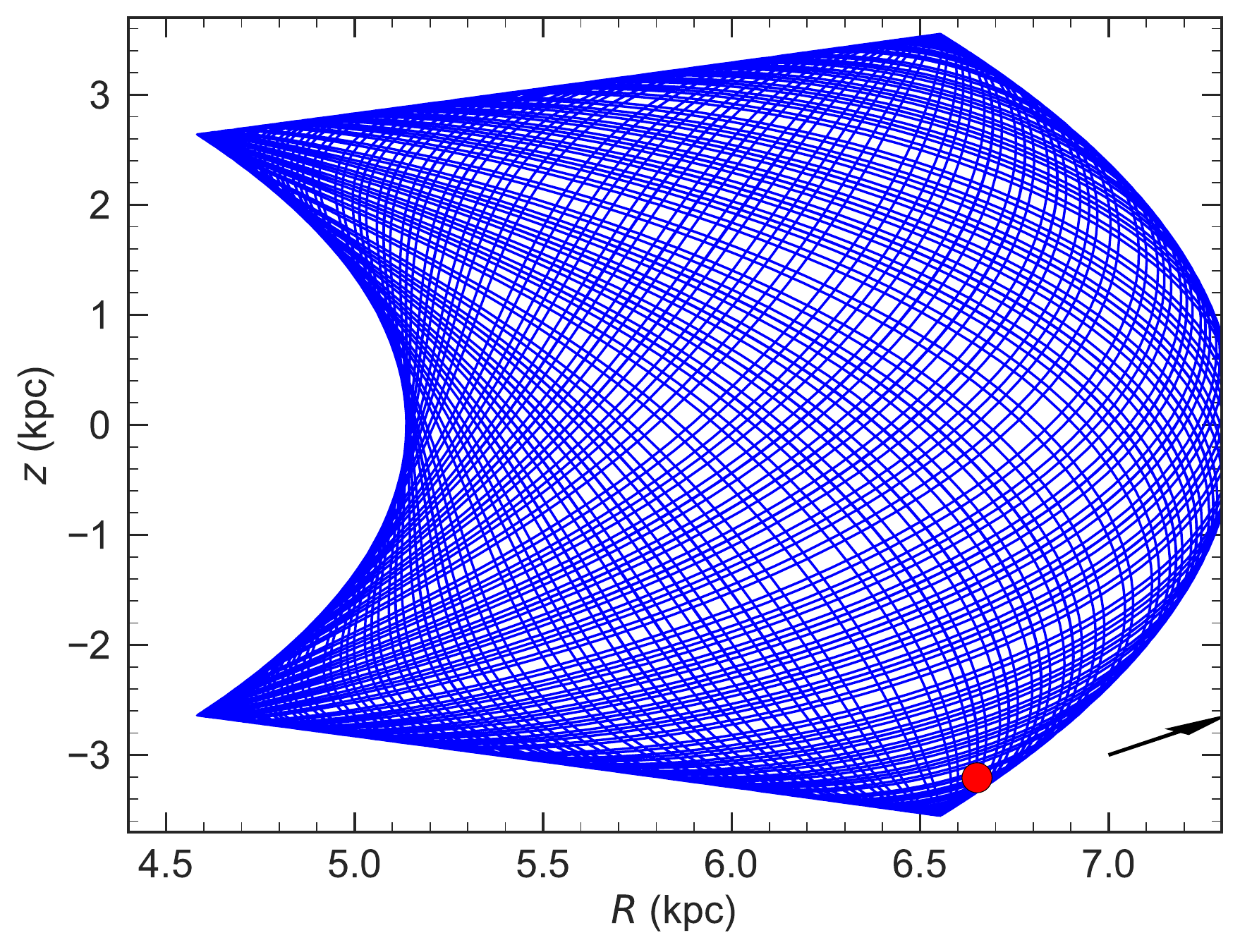}
\caption{Orbit of 47~Tuc in the Galactocentric $R$ vs $z$ plane. The orbit has been integrated back to 12 Gyr ago. The current position of the cluster is indicated by the red filled circle and the direction of the present day motion is represented by the black arrow.}
\label{fig:orbit}
\end{figure}

In this section, we will have a closer look at the orbit of 47~Tuc within the Milky Way. For this, we combine the cluster proper motion derived in this work with radial velocity measurements from the literature to get the full 3D velocity vector of the cluster. \citet{Marino16} performed a chemical abundance analysis of a sample of 75 RGB stars in 47~Tuc, thereby also deriving the radial velocity of the stars. They found a mean radial velocity of $-16.73 \pm 0.77~\mathrm{km~s^{-1}}$ with an $r.m.s.$ of 6.66~$\mathrm{km~s^{-1}}$. We will use this value for our analysis. For the spatial position of the cluster, we adopt the center of 47~Tuc as derived in \mbox{\citet{Li14}} ($\alpha_{2000} = 00^{\mathrm{h}}24^{\mathrm{m}}04\overset{\mathrm{s}}.80$, $\delta_{2000} = -72\degr04\arcmin48\arcsec$) and the distance of 4.57~kpc, resulting from our isochrone match to the data (see Section~\ref{sec:obs}). For the modeling of the cluster orbit, we made use of the python package \texttt{galpy}\footnote{\url{http://github.com/jobovy/galpy}} \citep{Bovy15}. We assumed a Milky Way-like potential, called \textit{MWPotential2014}, which is composed of three components, an exponentially cutoff power-law density bulge, a Miyamoto-Nagai disk and a dark matter halo as described in \citet{Navarro97}. The \texttt{galpy} package accepts the observed quantities of the velocity and position as input values and converts them to the corresponding values in the Galactic frame. We integrated the orbit backwards for 12~Gyr (approximately the age of the cluster). The orbit integration assumes a static potential which is not changing over the time of integration. This however, is just an approximation, since the Milky Way most likely has merged with a number of small galaxies in the past. For this reason, the orbit of the cluster becomes less reliable in the later stages of the simulation. Figure~\ref{fig:orbit} shows the resulting orbit in the Galactocentric $R$ vs $z$ plane, with the cluster's current position indicated as a red filled circle and the direction of the present day motion illustrated by the black arrow. The resulting ellipticity of the orbit is $e\sim$0.2. The pericenter and apocenter are at radii of $\sim$5.1~kpc and $\sim$7.5~kpc, respectively, whereas the maximum height above and below the Galactic plane is $\sim$3.6~kpc. According to our simulation, the cluster is confined within the inner regions of the Galaxy. 

Our results are in line with the findings of \citet{Odenkirchen97} and \citet{O'Malley17}. In their study, \citet{Odenkirchen97} derived the proper motions of a sample of 15 Galactic globular clusters from \textit{Hipparcos} data and combined their results with quantities from the \citet{Harris96} catalog to determine the overall motions of the clusters. For the calculation of the orbital parameters, they used a three-component potential. \citet{Odenkirchen97} found for 47~Tuc an orbit that extends between 4.3 and 7.9~kpc from the Galactic Center with a maximum height above the Galactic plane of 4.3~kpc. Additionally, they found an orbit ellipticity of 0.30. Recently, \citet{O'Malley17} analyzed the orbits of a sample of 22 clusters, including 47~Tuc. They used for the proper motions an average of the \citet{AndersonKing03} and \citet{Freire03} results and for the line-of-sight velocity, as well as the position and distance, the entries in the \citet{Harris96} catalog (2010 edition). \citet{O'Malley17} found 47~Tuc's orbit to be within 8~kpc from the Galactic Center with a maximum vertical height of 4~kpc and an ellipticity of 0.21 (see their Figure~4). The similarity of the results from these studies indicates that the orbital parameters are quite robust to moderate changes in the phase-space parameters of the cluster.

\section{Conclusions \label{sec:conclusions}}

In this paper we used multi-epoch observations from the VMC survey to derive the proper motions of stars within the globular cluster 47~Tuc. Taking advantage of the improved stellar position determinations from PSF photometry and an increased time baseline we found a median proper motion of the cluster of  $(\mu_{\alpha}\mathrm{cos}(\delta),~\mu_{\delta}) = (+5.89 \pm 0.02~\mathrm{(statistical)} \pm 0.13~\mathrm{(systematic)}, -2.14 \pm 0.02~\mathrm{(statistical)} \pm 0.11~\mathrm{(systematic)})$~mas~yr$^{-1}$. We also calculated the absolute proper motion of the cluster using data from the UCAC5 catalog. There we found $(\mu_{\alpha}\mathrm{cos}(\delta),~\mu_{\delta}) = (+5.30 \pm 0.03~\mathrm{(statistical)} \pm 0.70~\mathrm{(systematic)}, -2.70 \pm 0.03~\mathrm{(statistical)}\pm 0.70~\mathrm{(systematic)})$~mas~yr$^{-1}$, which agrees with the results from the VMC data within the uncertainties. Combining our results with radial velocity measurements from the literature, we were able to reconstruct the cluster's orbit for its current lifetime ($\sim$12~Gyr), showing that 47~Tuc was confined within the inner $\sim$7.5~kpc of the Milky way with a maximum vertical distance from the Galactic plane of about 3.6~kpc. 

When comparing our results for the proper motion with various previous determinations from the literature, we found that both the VMC values are closest to the measurements from \citet{Narloch17}, and \citet{AndersonKing03} (using updated proper motions for the SMC). However, our results differ from those presented by \citet{Cioni16}, using stellar positions from the VDFS pipeline, 
by 8$\sigma$ in $\mu_{\alpha}\mathrm{cos}(\delta)$ and 4$\sigma$ in $\mu_{\delta}$. 
We identified the longer time span between the observations as well as the differences in the stellar centers, especially in the most crowded regions, as the main sources that lead to this discrepancy. 
It is shown that the presented technique to derive proper motions using data from the VMC survey, with the above mentioned improvements, works very well for our test case of 47 Tuc. 

In upcoming studies we will use the wealth of VMC data and apply the techniques developed and described in this preparatory work to study the overall and internal kinematics of the SMC. The VMC survey is designed such that, on average, all tiles have a time baseline of $\sim$2 years for the multi-epoch $K_s$ observations. We will thereby concentrate on the SMC-Bridge region where \citet{Subramanian17} found large differences in the line-of-sight distance of the stars, but also on the kinematics of the different stellar populations that are shown to have large spatial differences within the galaxy \citep[e.g.][]{Ripepi17}. Adding a kinematic component to these observations will help us to better understand the history of the Magellanic System.  

\begin{acknowledgements}
We thank the Cambridge Astronomy Survey Unit (CASU) and the Wide Field Astronomy Unit (WFAU) in Edinburgh for providing calibrated data products under the support of the Science and Technology Facility Council (STFC). 

This project has received funding from the European Research Council (ERC) under European Union's Horizon 2020 research and innovation programme (grant agreement No 682115).

R. d. G. is grateful for research support from the National Natural Science Foundation of China (NSFC) through grants 11373010, 11633005, and U1631102.

This research made use of Astropy, a community-developed core Python package for Astronomy \citep{Astropy13}.

This study is based on observations obtained with VISTA at the Paranal Observatory under program ID 179.B-2003.

We thank the anonymous referee for useful comments and suggestions that helped to improve the paper.
\end{acknowledgements}

%
   \bibliographystyle{aa} 
   \bibliography{references} 
%

\begin{appendix}
\section{Proper Motion Trends}

In the Appendix, we present additional plots to look for any trends of the proper motion as a function of different quantities. In all figures the reflex proper motions of galaxies and the proper motions of the cluster stars is shown separately. For the galaxies, the reflex proper motion is calculated after the galaxy-based initial transformation and for the cluster stars, the proper motion is determined after the refined transformation to a common reference frame using the cluster stars itself. Therefore, it is expected that for both types of sources the proper motions have a mean value of zero (indicated by a dashed line in each panel).  

\begin{figure*}
 \begin{tabular}{cc}
  \includegraphics[width=0.9\columnwidth]{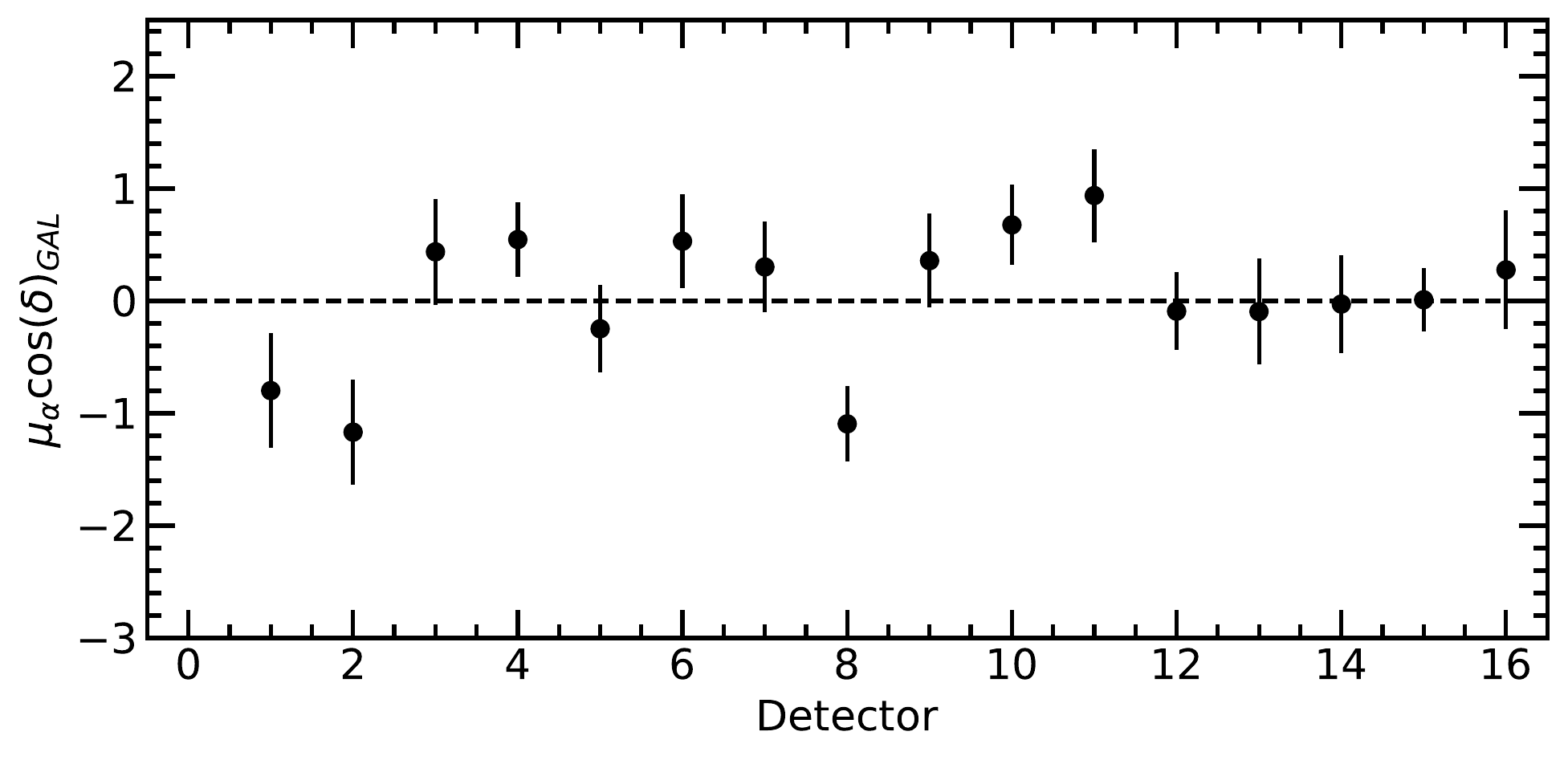} &
  \includegraphics[width=0.9\columnwidth]{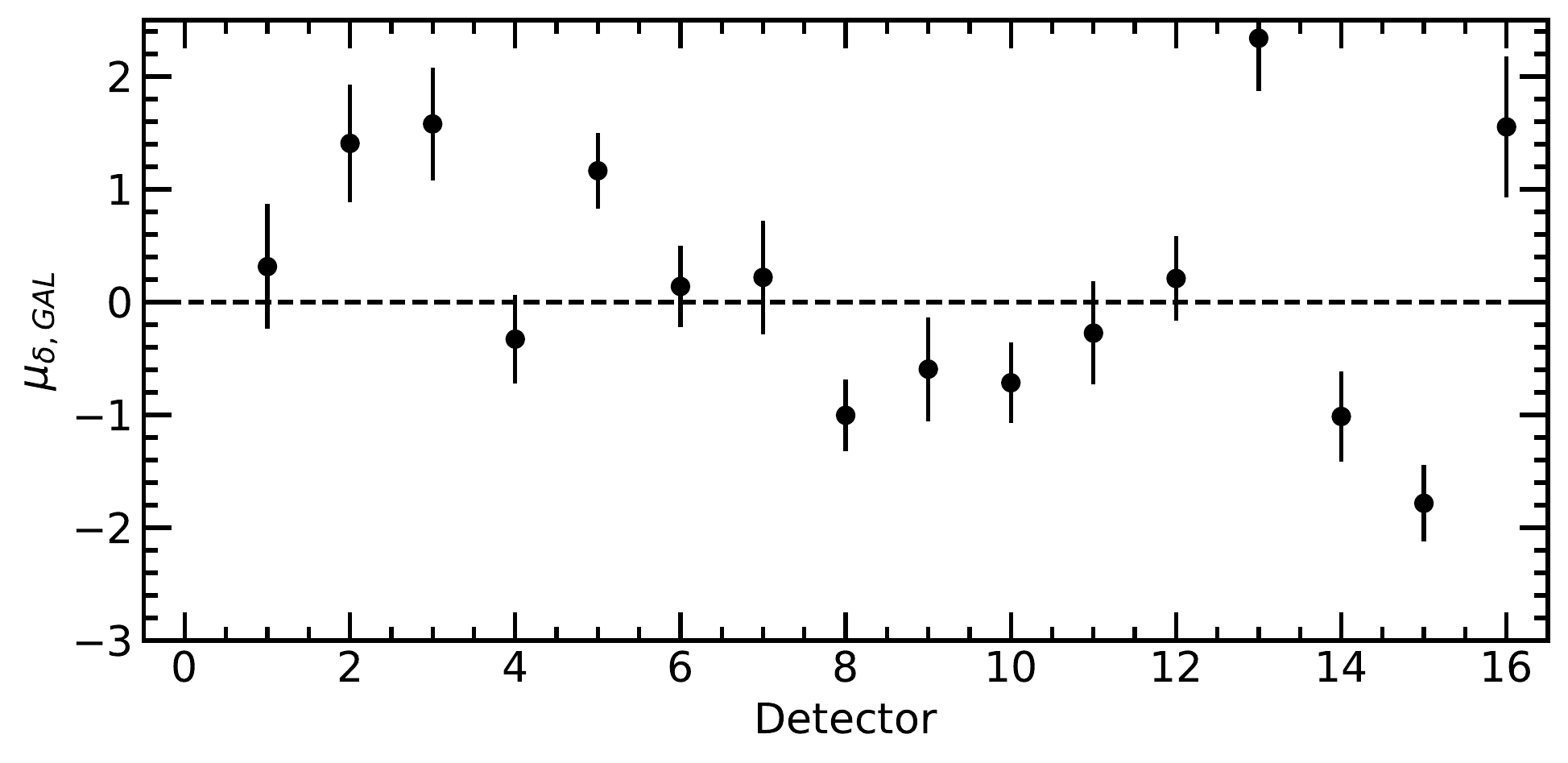} \\
    \includegraphics[width=0.9\columnwidth]{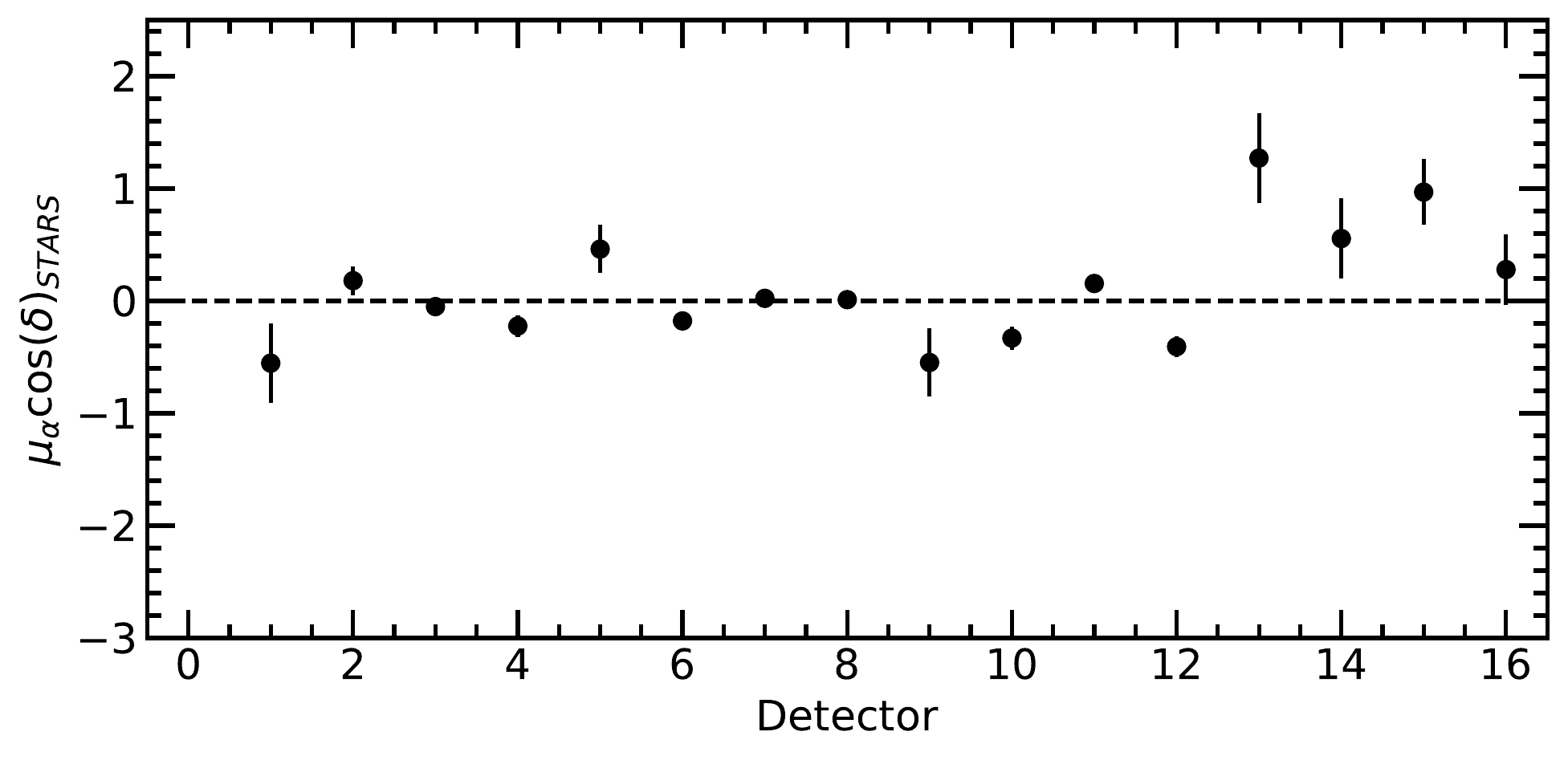} &
  \includegraphics[width=0.9\columnwidth]{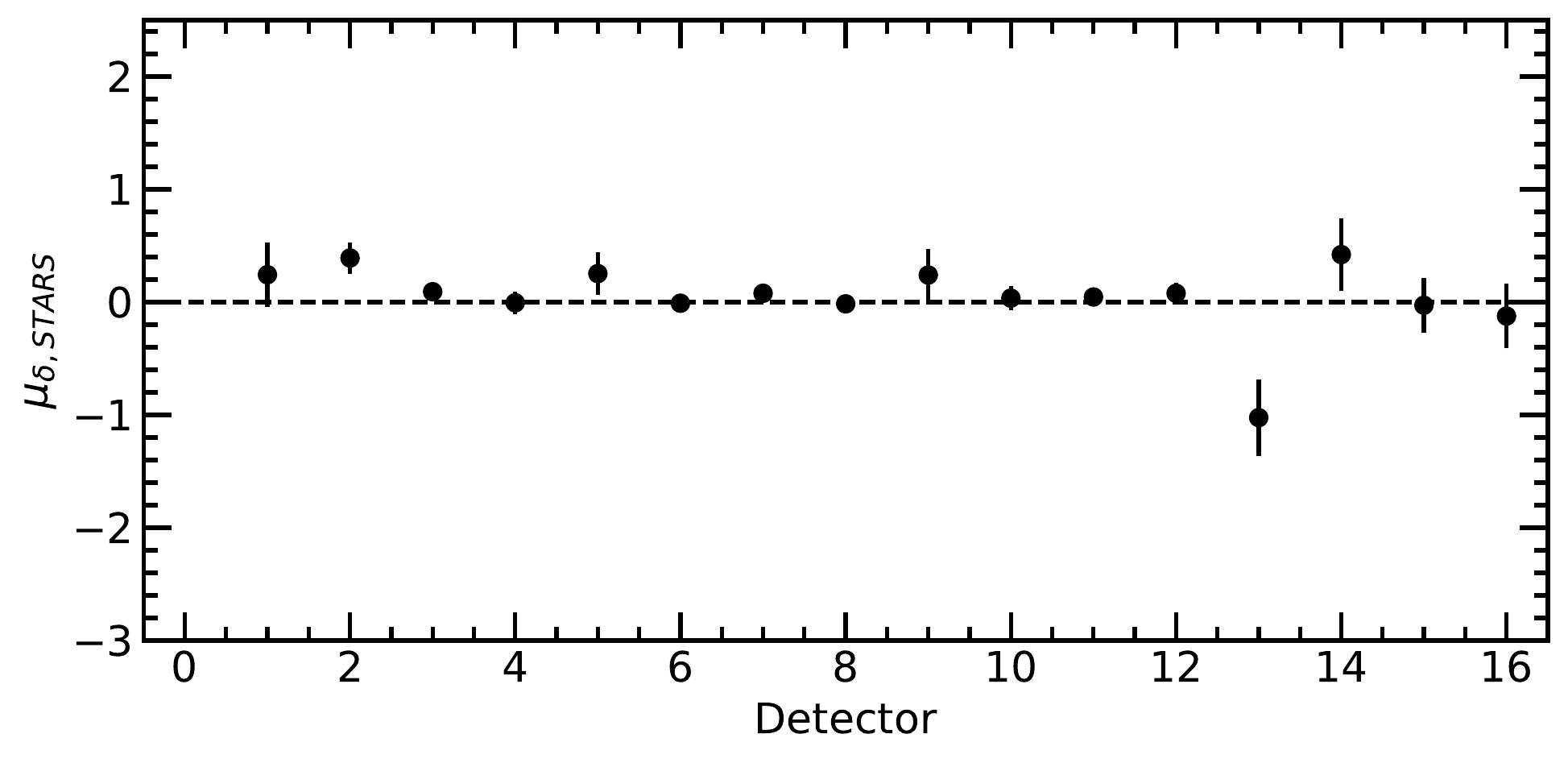} \\
 \end{tabular}
  \caption{\textbf{Top Panels:} Median reflex proper motions of galaxies in mas~yr$^{-1}$ on each detector. The error bars are the MAD divided by the square root of the number of galaxies within each detector. The left-hand panel shows the proper motions in the RA direction whereas the right-hand panel is for the Dec direction.  \textbf{Bottom Panels:} Same as the top panels but now for the proper motions of the cluster stars.}
   \label{fig:pm_vs_detector}
\end{figure*}

Figure~\ref{fig:pm_vs_detector} shows in the top panels the median reflex proper motions of the galaxies on each detector. The error bars indicate the median absolute deviation (MAD) divided by the square root of the number of galaxies within each detector. The bottom panels show the same for the proper motion of the cluster stars.

\begin{figure*}
 \begin{tabular}{cc}
  \includegraphics[width=0.9\columnwidth]{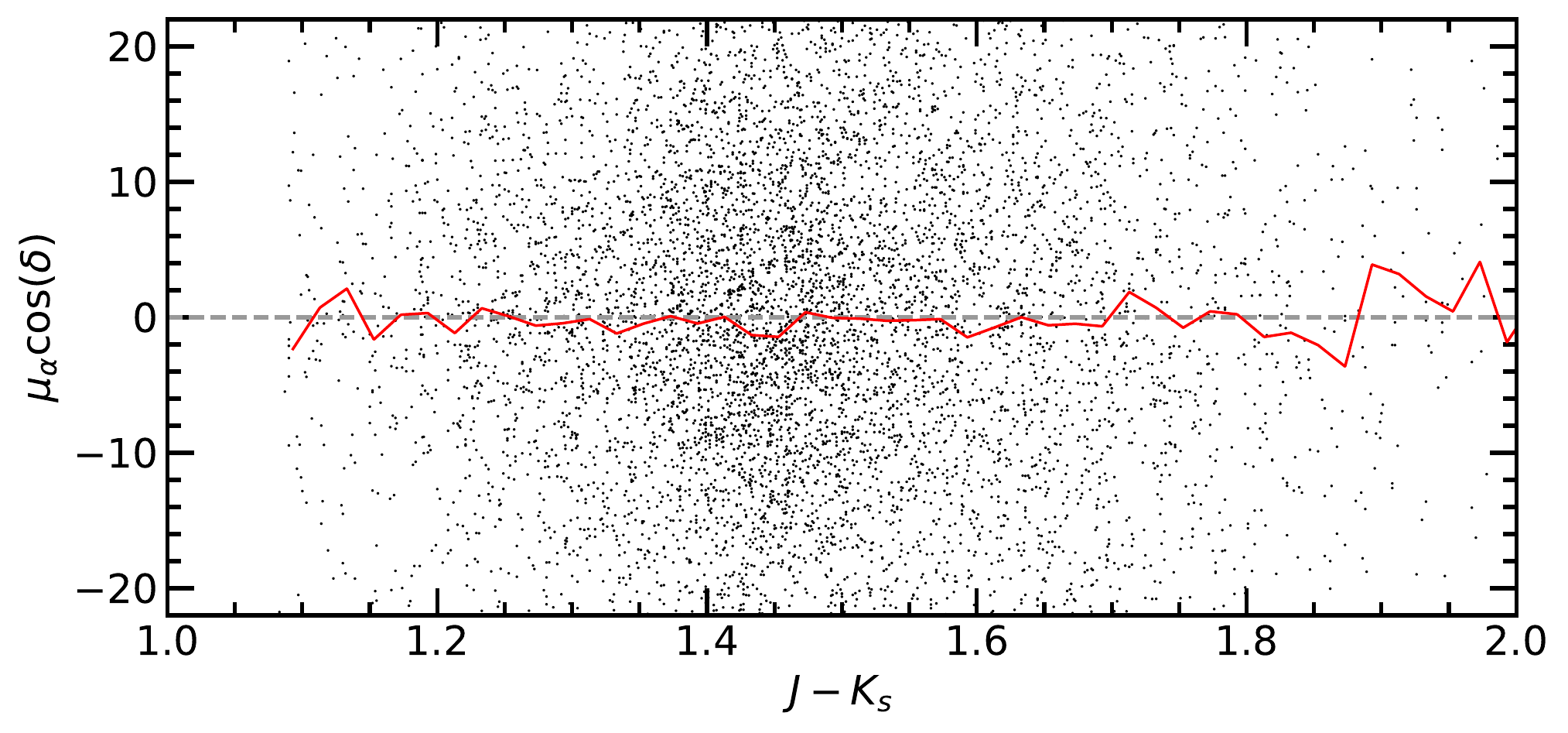} &
  \includegraphics[width=0.9\columnwidth]{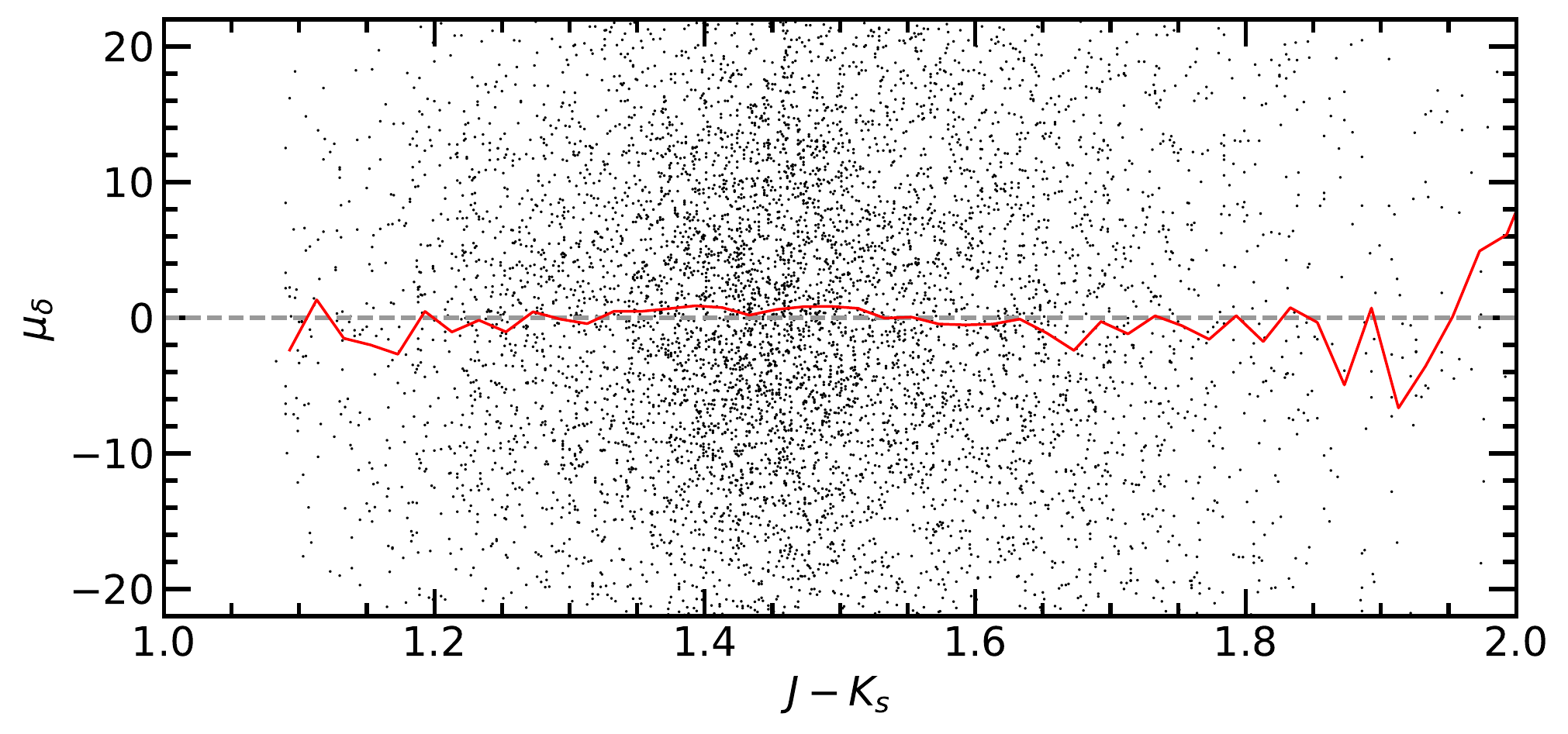} \\
    \includegraphics[width=0.9\columnwidth]{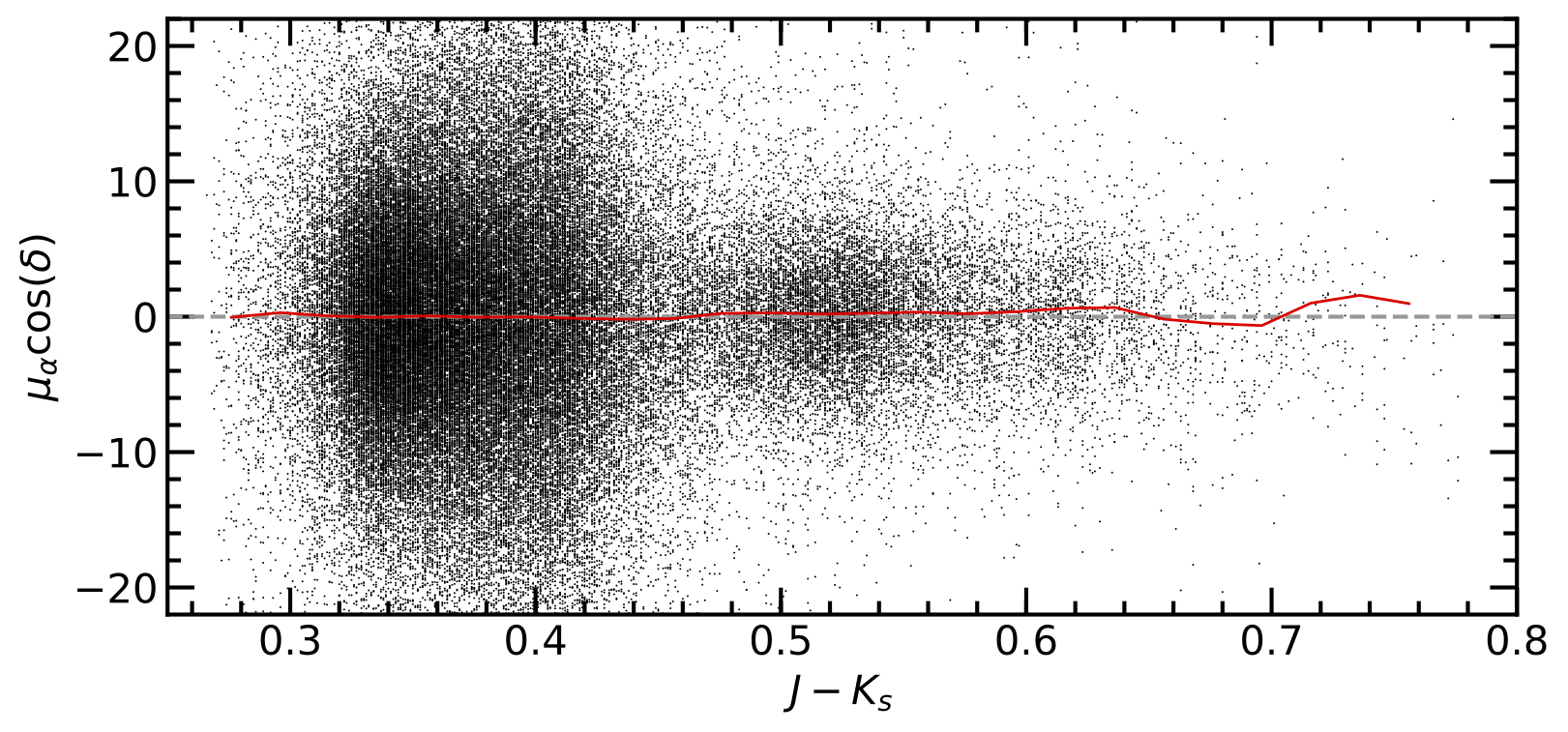} &
  \includegraphics[width=0.9\columnwidth]{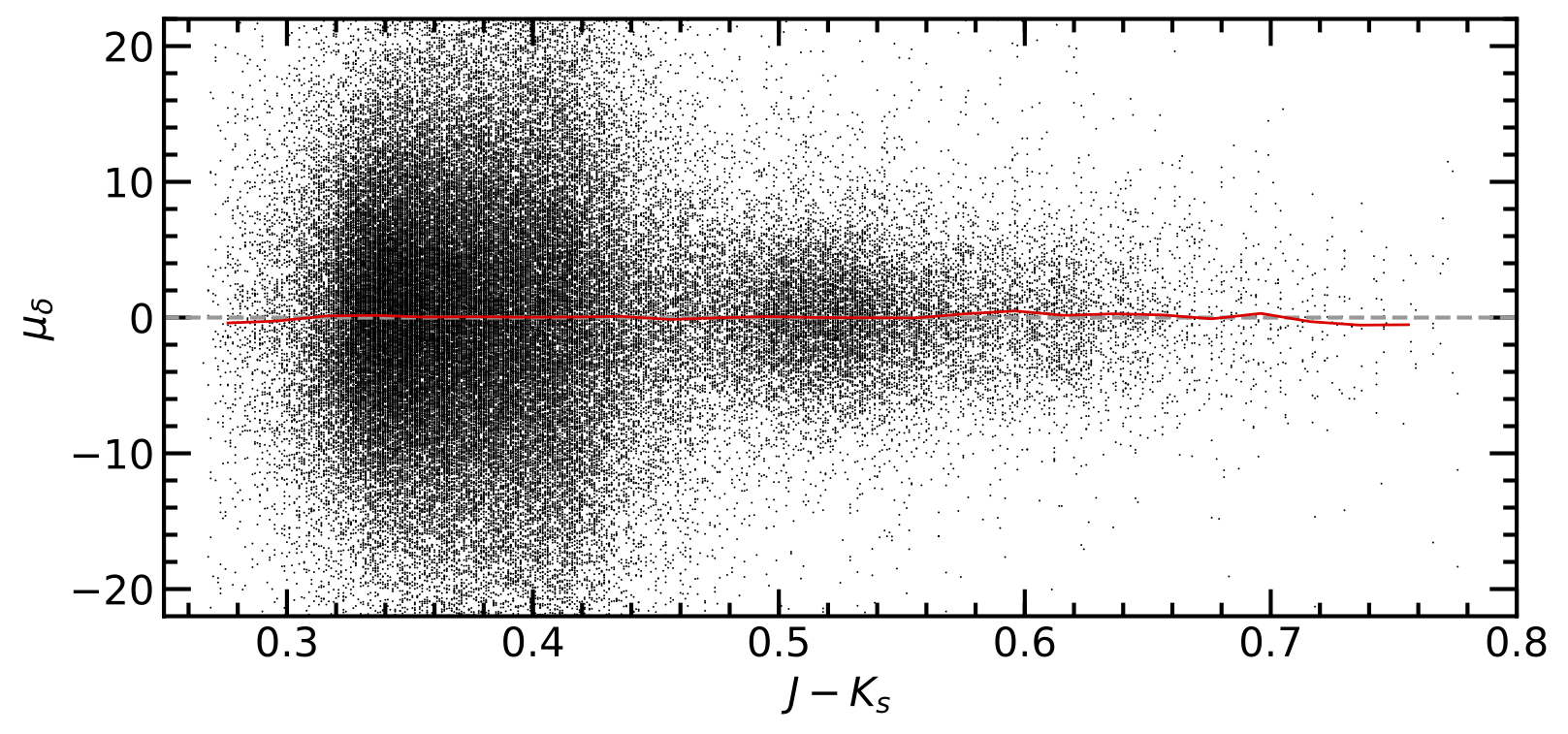} \\
 \end{tabular}
  \caption{\textbf{Top Panels:} Individual reflex proper motion values of galaxies in mas~yr$^{-1}$ as a function of the $J-K_s$ color. The running mean of this distribution is shown as the red solid line. The left-hand panel shows the proper motion in the RA direction whereas the right-hand panel is for the Dec direction.  \textbf{Bottom Panels:} Same as the top panels but now for the proper motions of the cluster stars.}
   \label{fig:pm_vs_colour}
\end{figure*}

The top panels of Figure~\ref{fig:pm_vs_colour} show the individual reflex proper motions in both directions of the galaxies as a function of the $J-K_s$ color. The red solid line indicates the running mean of all sources. The bottom panels show the same for the proper motions of the cluster stars. For both types of sources, no trends of the proper motions with color are found.

\begin{figure*}
 \begin{tabular}{cc}
  \includegraphics[width=0.9\columnwidth]{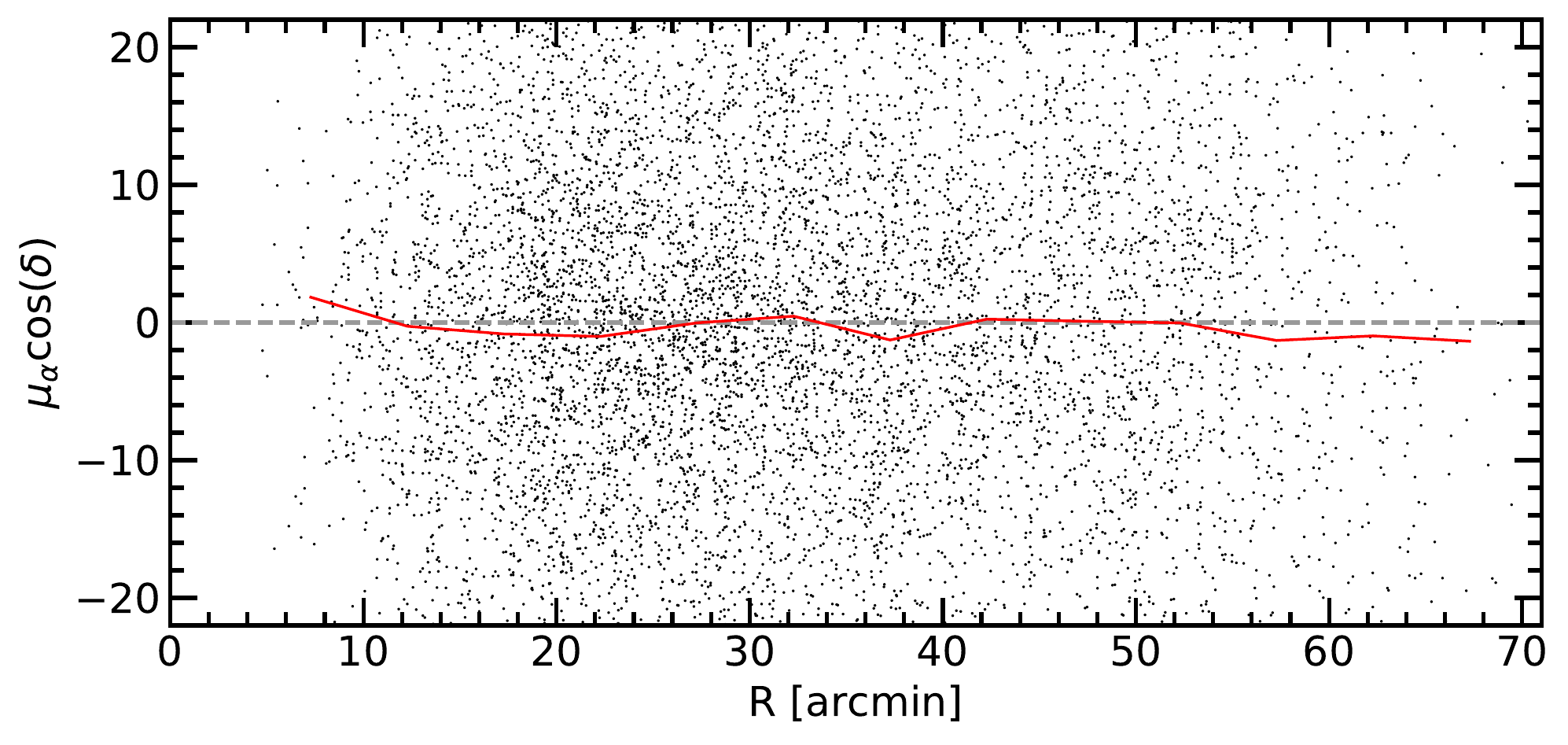} &
  \includegraphics[width=0.9\columnwidth]{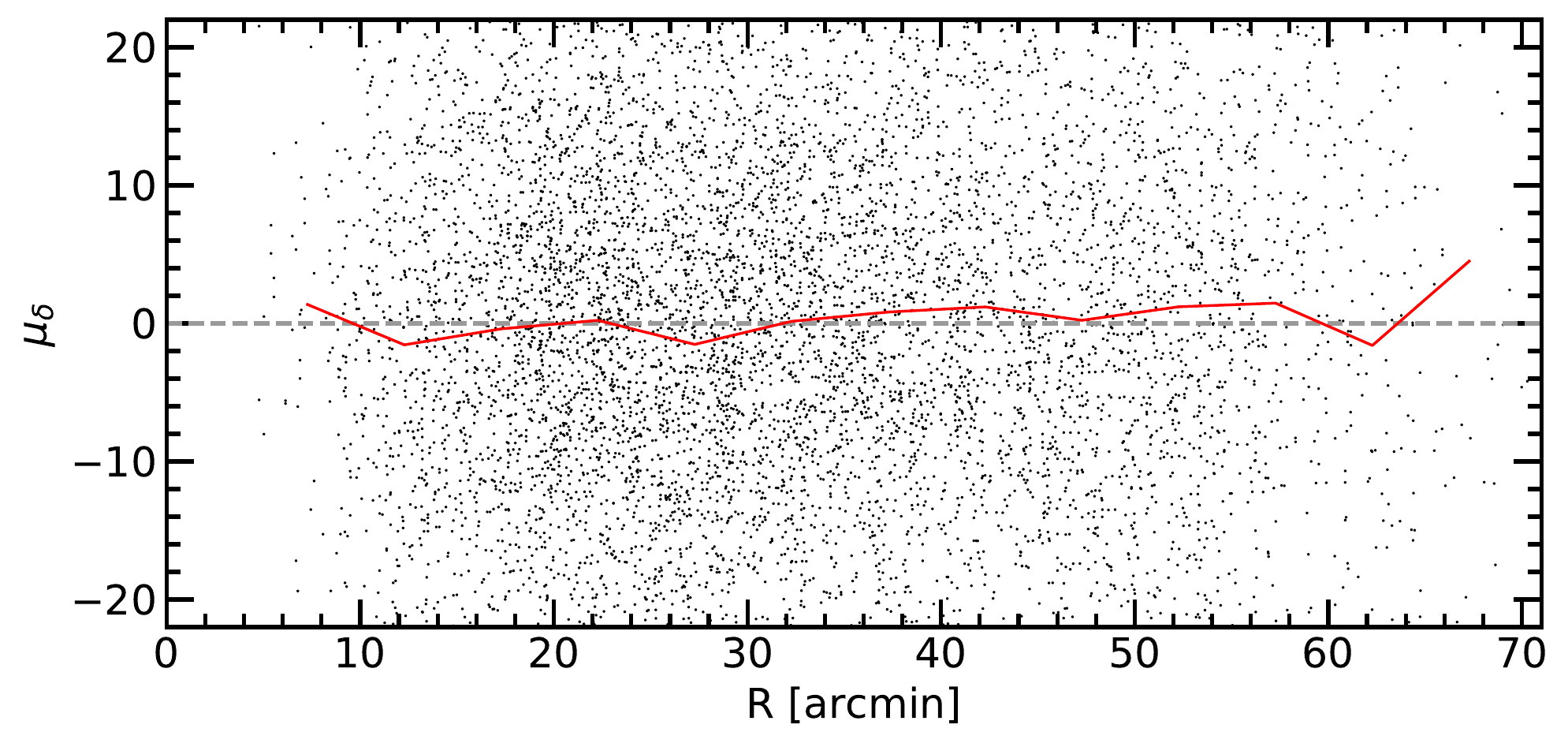} \\
    \includegraphics[width=0.9\columnwidth]{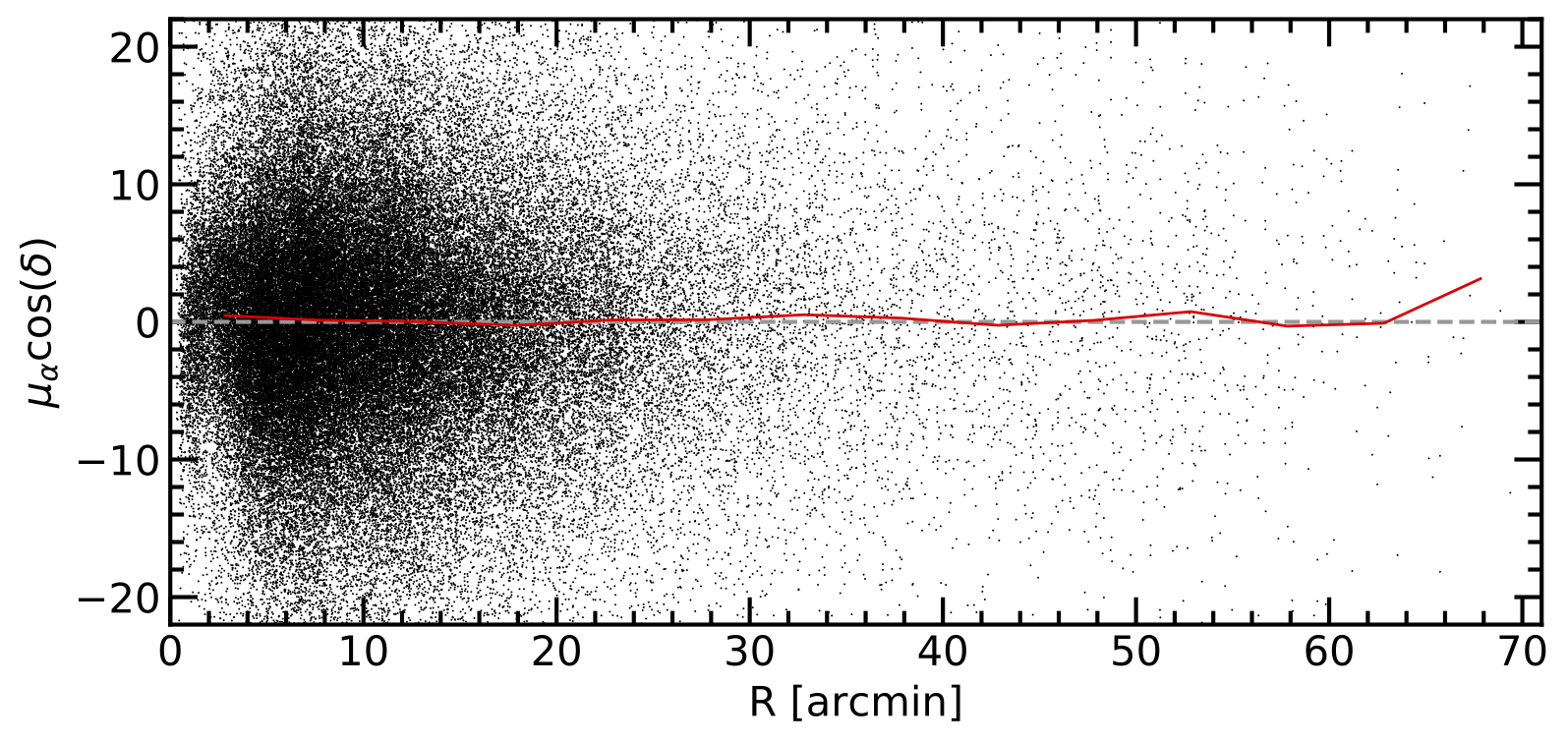} &
  \includegraphics[width=0.9\columnwidth]{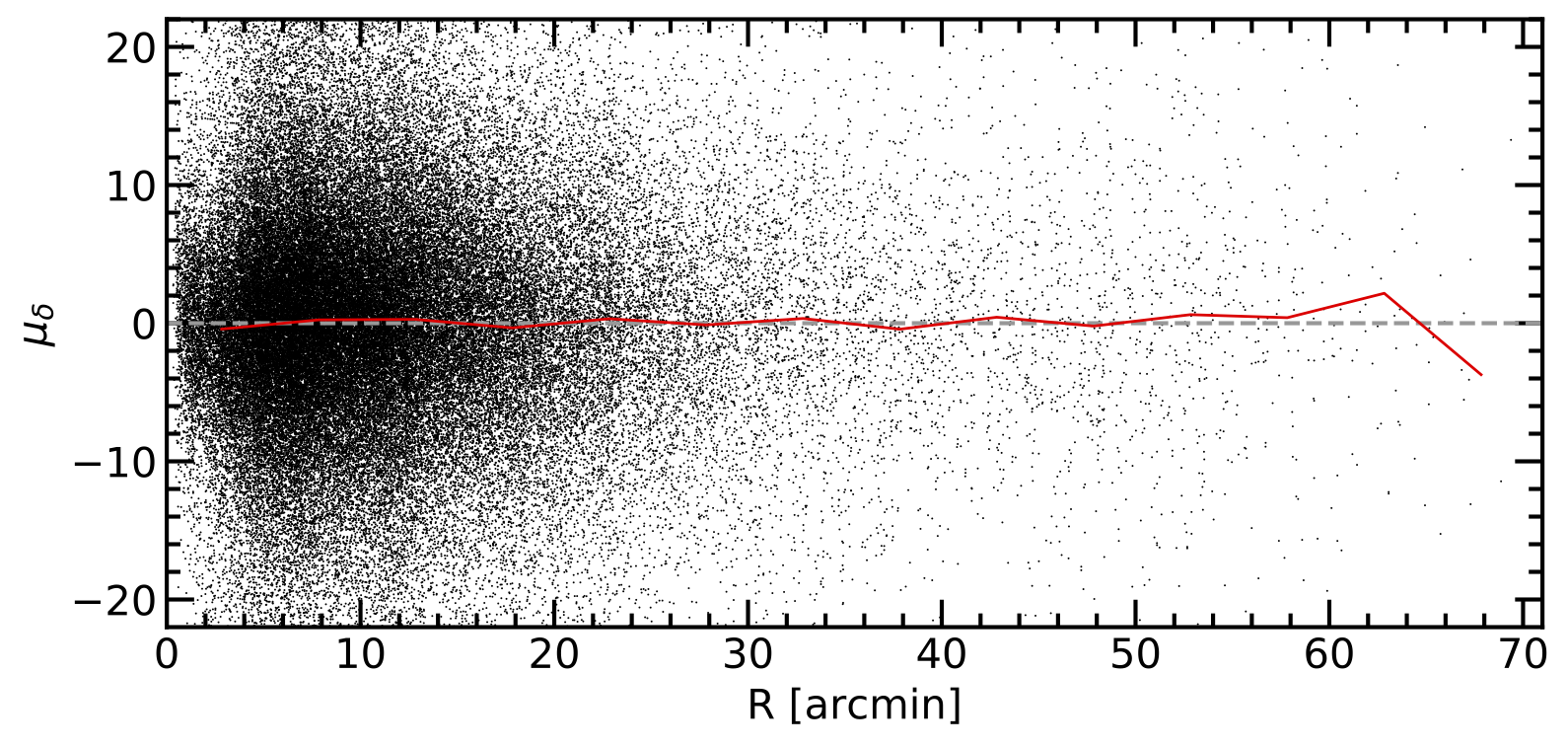} \\
 \end{tabular}
  \caption{\textbf{Top Panels:} Reflex proper motions of individual galaxies in mas~yr$^{-1}$ as a function of the distance in arcmin to the center of 47~Tuc. The red solid line marks the running mean of the distribution. The left-hand panel shows the proper motion in the RA direction whereas the right-hand panel is for the Dec direction. \textbf{Bottom Panels:} Same as the top panels but now for the proper motions of the cluster stars.}
   \label{fig:pm_vs_radius}
\end{figure*}

Figure~\ref{fig:pm_vs_radius} is similar to Figure~\ref{fig:pm_vs_colour} but now the proper motions as a function of the distance from the center of 47~Tuc are shown. There might be a slight trend of the galaxies' reflex motion as a function of the distance to the cluster center. In all other cases, there are no obvious trends seen in the proper motion.

\end{appendix}

\end{document}